\documentclass[longauth]{aa}  

\usepackage{graphicx}
\usepackage{xcolor}
\usepackage{txfonts}
\usepackage{lipsum}
\usepackage{soul}
\usepackage{subcaption}
\usepackage{lscape}
\usepackage{placeins}
\usepackage[colorlinks=true,citecolor=blue, linkcolor=blue, urlcolor=blue]{hyperref}

\newcommand\HD{HD\,121617}

\begin{document}

    \title{The ALMA survey to Resolve exoKuiper belt Substructures (ARKS)}

   \subtitle{IX: Gas-driven origin for the continuum arc in the debris disc of \HD{}.}

\author{P.~Weber\textsuperscript{1,2,3}\fnmsep\thanks{E-mail: philipppweber@gmail.com} \and S.~P\'erez\textsuperscript{1,2,3} \and C.~Baruteau\textsuperscript{4} \and S.~Marino\textsuperscript{5} \and F.~Castillo\textsuperscript{1,2,3} \and M.~R.~Jankovic\textsuperscript{6} \and T.~Pearce\textsuperscript{7} \and M.~C.~Wyatt\textsuperscript{8} \and A.~A.~Sefilian\textsuperscript{9} \and J.~Olofsson\textsuperscript{10} \and G.~Cataldi\textsuperscript{11,12} \and J.~B.~Lovell\textsuperscript{13} \and C.~del~Burgo\textsuperscript{14,15} \and A.~M.~Hughes\textsuperscript{16} \and S.~Mac~Manamon\textsuperscript{17} \and A.~Brennan\textsuperscript{17} \and L.~Matr\`a\textsuperscript{17} \and J.~Milli\textsuperscript{18} \and B.~Zawadzki\textsuperscript{16} \and E.~Chiang\textsuperscript{19} \and M.~A.~MacGregor\textsuperscript{20} \and D.~J.~Wilner\textsuperscript{13} \and M.~Bonduelle\textsuperscript{18} \and J.~M.~Carpenter\textsuperscript{21} \and Y.~Han\textsuperscript{22} \and \'A.~K\'osp\'al\textsuperscript{23,24,25} \and P.~Luppe\textsuperscript{17}} 

\institute{
Departamento de Física, Universidad de Santiago de Chile, Av. V\'ictor Jara 3493, Santiago, Chile \and
Millennium Nucleus on Young Exoplanets and their Moons (YEMS), Chile \and
Center for Interdisciplinary Research in Astrophysics Space Exploration (CIRAS), Universidad de Santiago, Chile \and
IRAP, Universit\'e de Toulouse, CNRS, CNES, F-31400 Toulouse, France \and
Department of Physics and Astronomy, University of Exeter, Stocker Road, Exeter EX4 4QL, UK \and
Institute of Physics Belgrade, University of Belgrade, Pregrevica 118, 11080 Belgrade, Serbia \and
Department of Physics, University of Warwick, Gibbet Hill Road, Coventry CV4 7AL, UK \and
Institute of Astronomy, University of Cambridge, Madingley Road, Cambridge CB3 0HA, UK \and
Department of Astronomy and Steward Observatory, The University of Arizona, 933 North Cherry Ave, Tucson, AZ, 85721, USA \and
European Southern Observatory, Karl-Schwarzschild-Strasse 2, 85748 Garching bei M\"unchen, Germany \and
National Astronomical Observatory of Japan, Osawa 2-21-1, Mitaka, Tokyo 181-8588, Japan \and
Department of Astronomy, Graduate School of Science, The University of Tokyo, Tokyo 113-0033, Japan \and
Center for Astrophysics | Harvard \& Smithsonian, 60 Garden St, Cambridge, MA 02138, USA \and
Instituto de Astrof\'isica de Canarias, Vía L\'actea S/N, La Laguna, E-38200, Tenerife, Spain \and
Departamento de Astrof\'isica, Universidad de La Laguna, La Laguna, E-38200, Tenerife, Spain \and
Department of Astronomy, Van Vleck Observatory, Wesleyan University, 96 Foss Hill Dr., Middletown, CT, 06459, USA \and
School of Physics, Trinity College Dublin, the University of Dublin, College Green, Dublin 2, Ireland \and
Univ. Grenoble Alpes, CNRS, IPAG, F-38000 Grenoble, France \and
Department of Astronomy, University of California, Berkeley, Berkeley, CA 94720-3411, USA \and
Department of Physics and Astronomy, Johns Hopkins University, 3400 N Charles Street, Baltimore, MD 21218, USA \and
Joint ALMA Observatory, Avenida Alonso de C\'ordova 3107, Vitacura 7630355, Santiago, Chile \and
Division of Geological and Planetary Sciences, California Institute of Technology, 1200 E. California Blvd., Pasadena, CA 91125, USA \and
Konkoly Observatory, HUN-REN Research Centre for Astronomy and Earth Sciences, MTA Centre of Excellence, Konkoly-Thege Mikl\'os \'ut 15-17, 1121 Budapest, Hungary \and
Institute of Physics and Astronomy, ELTE E\"otv\"os Lor\'and University, P\'azm\'any P\'eter s\'et\'any 1/A, 1117 Budapest, Hungary \and
Max-Planck-Insitut f\"ur Astronomie, K\"onigstuhl 17, 69117 Heidelberg, Germany
}

   \date{Received \today}

  \abstract
   {Debris discs were long considered to be largely gas-free environments, where dynamical evolution is governed primarily by collisional fragmentation, gravitational stirring, and radiative forces. 
   Recent detections of CO molecular line emission in debris discs demonstrate that gas is present, but its abundance and origin are still uncertain.
   The {\it ALMA survey to Resolve exoKuiper belt Substructures (ARKS)} observed both the gas and dust of several debris discs at high resolution and revealed a narrow ring of gas and dust in the disc \HD{}, with an asymmetric arc-like feature that is 40\% brighter than the rest of the ring.}
   {An important open question is how representative the estimated CO masses are for the total gas mass in debris discs.
   We aim to constrain the total gas mass in \HD{} using numerical models under the assumption that the dust arc is produced by hydrodynamical processes involving the gas.} 
   {We used the hydrodynamical code Dusty FARGO-ADSG, in which dust is modelled as Lagrangian particles.
   We explored the effects of radiation pressure and dust feedback, as well as of varying the total gas mass on the dynamical evolution of the system.
   We compared these simulations with observations via radiative transfer calculations.}
   {We find that an unstable gas ring can create a size-dependent radial and azimuthal dust trap. 
   The total gas mass dictates the efficiency of particle trapping as a function of grain size.
   We find that two of our models, $M_{\rm gas}=50\,M_\oplus$ and $M_{\rm gas}=5\,M_\oplus$, can simultaneously reproduce the observed arc in the ALMA band\,7 continuum image and the radial outward offset of the VLT/SPHERE scattered light ring, driven by the combined effects of gas drag and radiation pressure.
   We further find a conservative lower limit of $M_{\rm gas}>2.5\,M_\oplus$ and a conservative upper limit of $M_{\rm gas}<250\,M_\oplus$.}
   {If the ALMA band\,7 asymmetry is caused by gas drag, reconciling the required gas mass with the observed $^{12}$CO emission suggests the presence of significant amounts of H$_2$, consistent with the gas being primordial, that is, long-lived remnant material from the protoplanetary disc phase.
   In this scenario, \HD{} would represent a hybrid disc, bridging the protoplanetary and debris disc stages.
   As an arc-shaped emission can alternatively be reproduced by a planet’s gravitational forcing, future observations are crucial to distinguish between these two scenarios.
   }

   \maketitle

\nolinenumbers
\section{Introduction}
\label{sec:intro}
Debris discs consist of a dust population that is predominantly generated by collisions between larger bodies \citep[e.g.,][]{Hughes2018, Marino2022,Pearce2024}.
Therefore, they are generally regarded as second-generation discs in which dust is continually replenished.
The advent of the Atacama Large Millimetre/Submillimetre Array (ALMA) has not only provided unprecedented angular resolution to investigate thermal emission from these structures, but has also revealed that some debris discs harbour substantial amounts of gaseous carbon monoxide \citep[CO, e.g.][]{Kospal2013,Hughes2018,Cataldi2023} with mass estimates ranging from $10^{-6}\,M_\oplus$ to $10^{-1}\,M_\oplus$ \citep{Cataldi2023,gas_arks}.
In the pre-ALMA era, the prevailing view \citep[apart from the case of 49 Ceti,][]{Zuckerman1995} was that debris discs should be gas-free environments, since the primordial gas in protoplanetary discs is expected to be dispersed near the end of the disc phase by a combination of processes, including photoevaporation, magnetically driven (magnetothermal) winds, and viscous draining \citep[e.g.][]{Clarke2001,Alexander2014,Ercolano2017,Bai2016}.
Therefore, the detection of CO is of utmost interest, as it challenges this classical notion of gas-free debris discs.

In several systems, the estimated CO mass is too large to be consistent with the expected gas production rate from cometary mass loss \citep{Kral2017}, prompting questions about the origin of the CO.
Two favoured scenarios have been proposed: either the gas from the protoplanetary phase persists for longer than previously thought \citep[primordial gas, e.g.][]{Kospal2013,Nakatani2021}, or it is released during the debris disc phase from the volatile-rich solid bodies, either by collisions \citep[e.g.][]{Czechowski2007, Bonsor2023} or photodesorption \citep[][]{Grigorieva2007} and subsequently shielded from ultraviolet (UV) radiation \citep[secondary gas;][]{Kral2019,Marino2020}.

The chemical composition in the primordial and secondary scenarios differs considerably \citep[e.g.][]{Klusmeyer2021, Smirnov2022}.
In the secondary gas scenario, CO, C and O are the principal constituents; in contrast, the primordial-gas scenario implies a total gas mass dominated by molecular hydrogen, H$_2$, leading to a total gas mass significantly larger than that inferred from CO observations.
Although direct observation of gas tracers often struggles to distinguish between these scenarios, each has markedly different implications for dust dynamics.
Gas and dust co-located in the disc exchange momentum through friction, whose strength depends on both the grain surface area per unit mass and the local gas density \citep[e.g.][]{Safronov1972,Whipple1972}.
Consequently, a sufficiently dense gas phase is needed to exert a notable drag force on dust grains.
If composed of second-generation gas, the total gas mass would be relatively low and frictional coupling with millimetre‐sized dust would remain negligible \citep{Marino2020}.
In contrast, if the gas composition resembles that of typical protoplanetary discs (i.e. abundant, yet undetected in H$_2$), even the mm dust would likely experience significant dust–gas interactions.

While the interaction between dust and gas has been extensively investigated in protoplanetary discs \citep[see][for a recent review]{Lesur2023}, relatively few studies have focused on this process in the context of debris discs.
One of the main differences between the two environments is that debris discs are optically thin to stellar radiation, and studies of optically thick protoplanetary discs cannot be directly applied.
In optically thin discs, radiation pressure works in conjunction with gas drag to reshape the spatial distribution of small dust grains \citep{Takeuchi2001}.
Resulting effects on the dust radial and vertical structure have often been studied assuming a fixed gas distribution and an azimuthally symmetric disc \citep{Thebault2005,Krivov2009,Olofsson2022,hd131835_arks}, although there are several studies that considered frictional coupling between the dust and the gas \citep[e.g.][]{Klahr2005,Besla2007,Lyra2013,Castrejon2019}.
In studies focusing on gas production, destruction, and evolution of secondary gas \citep[e.g.][]{Kral2016,Kral2017,Kral2019,Hales2019,Marino2020}, or the effect of a planet on secondary gas dynamics \citep{Bergez-Casalou2024}, dust was treated only as a tracer species or omitted altogether.
Overall, these works highlight that hydrodynamical processes, previously assumed to be negligible in debris discs, may indeed be essential to understanding observed structures.

\begin{figure*}
    \centering
    \includegraphics[width=0.8\textwidth]{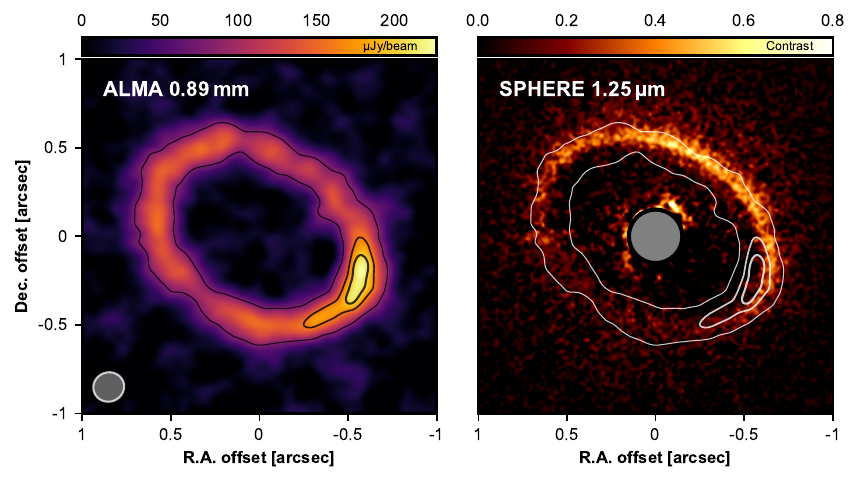}
    \caption{Observations of \HD{}. The left panel shows the ALMA band\,7 continuum observation ($\lambda_{\rm obs}=0.89\,$mm) from \citet{overview_arks}. The contours at $[80,160,200]\,\mu{\rm Jy}/{\rm beam}$ highlight the ring's azimuthal asymmetry. The clean beam is displayed as a white ellipse in the bottom left. The right panel shows the SPHERE/IRDIS $J$-band  scattered light image ($\lambda_{\rm obs}=1.25\,\mu$m) from \citet{scat_arks}, with the same ALMA continuum contours. The grey circle indicates the coronagraphic mask of the observation.}
    \label{fig:obs}
\end{figure*}
Within the sample of objects observed by the ALMA survey to Resolve exoKuiper belt Substructures \citep[ARKS,][]{overview_arks}, the debris disc around \HD{} stands out as a specifically interesting case.
The left panel of Fig.~\ref{fig:obs} shows the ARKS ALMA band\,6 observation of \HD{} \citep[][]{overview_arks} that exhibits an arc to the southwest in the dust continuum \citep[][]{hd121617_arks,asym_arks} reminiscent of protoplanetary disc structures formed through dust trapping in a gas vortex \citep[e.g.][]{Meheut2012,Baruteau2019}.
Comparable examples in protoplanetary discs include the systems MWC\,758 \citep[][]{Baruteau2019}, HD\,143006 \citep[][]{Perez_Laura2018} and PDS\,70 (\citealp{Doi2024}, Daza et al. in prep.).
The right panel of Fig.~\ref{fig:obs} shows the VLT/SPHERE $J$-band image of \HD{} presented in \citet{scat_arks} with the original data presented in \citet{Perrot2023}. The comparison shows that, in addition to its pronounced azimuthal asymmetry at mm wavelengths, the ring around \HD{} shows a radial outward offset in the $J$-band scattered light image relative to the radial peak of mm emission.
This is indicative of radial segregation between dust grains of different sizes, and a small eccentricity, but no significant asymmetry \citep{scat_arks}.
The combination of azimuthally confined continuum emission \citep[][]{asym_arks,hd121617_arks}, substantial CO gas content \citep[][]{Moor2017,gas_arks}, lack of an arc in the distribution of small grains, and the observed radial offset in scattered light \citep[][]{scat_arks} makes \HD{} a compelling target for investigating whether, and under which conditions, gas drag can shape the dust distribution in a debris disc.

\HD{} is an A1-type pre-main-sequence star located at a distance of 117.9$\pm0.4\,$pc \citep{GaiaDR3}, with an estimated age of $16\pm2\,$Myr and a luminosity of 14.0$^{+0.4}_{-0.3}\,$L$_\odot$\citep{Matra2025}, a mass of $1.901\pm0.009$\,$M_\odot$, a radius of $1.553\pm0.01\,R_\odot$, and an effective temperature of $9029\pm34\,$K \citep{overview_arks}.
\HD{}'s age places it as an object of interest possibly transitioning from the protoplanetary disc phase to a debris disc stage, marking it as a potential hybrid system with primordial gas.
The dust mass is estimated to be $0.21\pm0.02\,M_\oplus$ (\citealp{overview_arks}, see also \citealp{Perrot2023}), while radiative transfer models of \HD{} show that the molecular line emission from $^{13}$CO and $^{12}$CO are optically thick indicating (model dependent) masses as high as $M\left(^{13}\mathrm{ CO}\right){=}2^{+0.5}_{-0.7}\,{\times}10^{-3}\,M_\oplus$ and $M\left(^{12}\mathrm{ CO}\right){\sim}0.15\,M_\oplus$, respectively \citep{line_arks}.
If the disc retains a significant fraction of primordial gas inherited from the protoplanetary disc stage, the total gas mass would likely exceed the measured CO mass by several orders of magnitude \citep[see][for a discussion]{Williams2014}.  
In particular, \HD{} exhibits several signatures that we investigate in the context of dust–gas interactions in a debris disc.
The key observables are:
(i) an azimuthally confined dust overdensity producing a local brightness enhancement of about 40\% at 0.89\,mm, with an azimuthal extent characterised by a full width at half maximum (FWHM) of roughly 90$^\circ$ \citep{hd121617_arks};
(ii) a near-infrared scattered-light ring located about 9.5\% farther out than the submillimetre ring \citep[][]{scat_arks}; and
(iii) a mirror symmetry of the scattered-light ring with respect to the semi-minor axis \citep{scat_arks}.

In this paper, we present a model that qualitatively explains these features as the outcome of dust–gas interactions.
Our results show that an unstable gas ring can simultaneously explain the azimuthal brightness variations seen in the ALMA band\,7 continuum and the radial offset of the scattered light ring.
At the outset, however, we note that our proposed explanation for the observations of \HD{} is not unique; a gas-free scenario in which gravitational forcing by a migrating planet produces the observed structure will be explored in future work (Pearce et al., {\em in preparation}).

The rest of the paper is structured as follows: We present the simulation setup and numerical methods in Sec.~\ref{sec:hydro}.
We present the main results in Sec.~\ref{sec:results}.
An observational comparison based on radiative transfer and image synthesis is presented in Sec.~\ref{sec:RT}.
We discuss our findings and suggest tests for future observations in Sec.~\ref{sec:discussion} and conclude in Sec.~\ref{sec:conclusions}. Technical details of calculations are presented in Appendices \ref{appendix:feedback} and \ref{appendix:timeseries}. 

\section{Hydrodynamical simulations}\label{sec:hydro}
Hydrodynamical simulations are commonly used to explore protoplanetary disc dynamics, where it is generally assumed that the total disc mass is dominated by gas.
On the other hand, hydrodynamics had long been considered irrelevant for the study of debris discs, until the discovery of significant CO emission in various exoKuiper belts raised the possibility that gas could affect dust structures and evolution.
A key distinction between protoplanetary discs and debris discs is the difference in both dust and gas densities.
The higher dust masses of protoplanetary discs generally make them optically thick to the stellar irradiation, and their gas densities are sufficiently high for micron-sized grains to remain tightly coupled to the gas flow.
By contrast, debris discs are commonly believed to be optically thin at all wavelengths (inferred from their small fractional luminosities), making small dust grains susceptible to radiation pressure.
Furthermore, the lower gas density in debris discs (especially for the secondary gas origin) implies that even micron-sized grains decouple from the gas dynamics.
While an interstellar medium–like (ISM-like) dust-to-gas ratio of about 1\% is typically assumed in protoplanetary discs \citep{Williams2011}, this ratio can be much larger in debris discs \citep[e.g.][]{Moor2017}.

These differences in gas density imply that not all techniques developed for protoplanetary disc simulations can be directly applied to debris discs.
For example, the low gas densities expected in debris discs imply long momentum stopping times for the dust.
This opens up the possibility that multiple streams of dust moving in multiple directions can be present at a single location, making it inappropriate to treat dust as a pressureless fluid with a single velocity at a given location \citep[as already highlighted in a simplistic approach by][]{Lynch2024}.
Meanwhile, stellar irradiation exerts a significant radiation pressure on dust particles throughout the disc that can become relevant for small dust grains of a high surface-to-mass ratio, necessitating an explicit numerical treatment of this force, as done here.
Finally, because of the potentially high dust-to-gas ratio in debris discs, we include both the frictional effect of the gas onto the dust and the back-reaction of the dust onto the gas dynamics.

We set up numerical simulations to determine whether gas friction influences dust structures around \HD{}, specifically the arc observed in the ALMA band\,7 continuum \citep{overview_arks} and the offset seen in scattered light \citep{scat_arks}.
We used the hydrodynamical code Dusty FARGO-ADSG\footnote{\url{https://github.com/charango/dustyfargoadsg}} \citep{Baruteau2016}, which is based on the FARGO algorithm \citep[a fast Eulerian transport algorithm for differentially rotating discs,][]{Masset2000}, and allows gas and dust to evolve in two dimensions (radial and azimuthal).
In this framework, the gas is defined as an Eulerian fluid on a fixed grid, while the dust is represented by Lagrangian particles, i.e. discrete fluid elements that are tracked as they move through the flow field.
Here, each computational particle (superparticle) in the simulation can be thought of as a collection of physical dust grains.
The following subsections describe how we initialise the key hydrodynamic fields and outline essential numerical details.

\subsection{Disc setup}
We initialised the gas density as an azimuthally smooth ring around the central star, represented by a Gaussian radial profile atop a Gaussian background function, motivated by the radial profiles presented in \citet{rad_arks}.
Specifically, we set the initial gas surface density to
\begin{equation}\label{equ:dens0}
\Sigma_{\rm g} = \Sigma_{\rm 0} \times \exp\left(-\frac{(r-r_0)^2}{2\sigma^2}\right)+\Sigma_{\rm bg}\times\exp\left(-\frac{(r-r_{\rm bg})^2}{2\sigma_{\rm bg}^2}\right),
\end{equation}
where $r_0$ and $r_{\rm bg}$ are the characteristic radii of the ring centre and the background centre, respectively, and $\sigma$ and $\sigma_{\rm bg}$ are the ring’s and the background's half-widths.
The ring’s stability is particularly sensitive to its width $\sigma$ and to the contrast between the ring's maximum density $\Sigma_0$ and the background density \citep[e.g.][]{Ono2018}.
In our simulations, the initial density distribution corresponded to a narrow ring (small $\sigma$) with a high contrast, defined as $\mathcal{C} \equiv \Sigma_0 / \Sigma_{\rm bg} \gg 1$. In Section~\ref{subsec:2D}, we chose these parameters such that the ring became susceptible to the Rossby Wave Instability (RWI; \citealt{Lovelace1999}).
The RWI can be triggered when the disc’s vortensity (the ratio of flow vorticity to density) exhibits a local extremum \citep[necessary but not sufficient criterion,][]{Lovelace1999}\footnote{A sufficient criterion has been found empirically in \citet{Chang2023}.}.

Even in protoplanetary discs, where gas dynamics has long been seen as the most important driver of evolution, the magnitude and origin of viscosity remain poorly constrained \citep[see][for comprehensive reviews]{Lesur2023,Rosotti2023}.
This uncertainty is even greater in debris discs, where the relevance of gas dynamics has only been considered in recent years.
For a comprehensive review of potential viscosity sources in these environments, we refer to \citet{Cui2024}.
Here, we employed the $\alpha$-viscosity model \citep{Shakura1973} for the kinematic viscosity,
$\nu = \alpha\,c_{\mathrm{s}}^{2}\,\Omega_{\mathrm{K}}^{-1}$, where $c_\mathrm{s}$ is the sound speed and $\Omega_\mathrm{K}$ the Keplerian angular frequency.
We assumed a locally isothermal equation of state and adopted a gas temperature profile of $T\sim r^{-0.5}$, a standard choice for debris discs \citep[e.g.][]{Kral2017}.
From this, we obtained the pressure scale height $H=c_{\rm s} \Omega_{\rm K}^{-1} \sim r^{1.25}$ and the aspect ratio $h=H/r \sim r^{0.25}.$ 
We emphasise that this flaring index of 0.25 remains open to debate, as it is often motivated by a black-body dust profile \citep[][]{Backman1993}. Different heating and cooling mechanisms for dust and gas mean that they are not necessarily thermally coupled \citep[e.g.][]{Zagorovsky2010}; in fact, in \HD{}, gas and dust temperatures were assumed to be partially different \citep{line_arks}.

To account for non-Keplerian rotation due to the radial pressure gradient, we modified the initial azimuthal gas velocity as
\begin{equation}\label{equ:vphi}
u_\phi = v_{\rm K}\sqrt{1 - \eta}\,; \qquad \eta \equiv - h^2\frac{d\ln P}{d\ln r}\,,
\end{equation}
where $v_{\rm K}$ is the Keplerian velocity, and $P=\Sigma_{\rm g}c_{\rm s}^2$ is the vertically integrated locally isothermal gas pressure.
The quantity $\eta$, therefore, characterises how much the azimuthal velocity of the gas deviates from Keplerian rotation.
Conceptually, if the pressure decreases with radius, $\eta$ will be positive and the gas will, therefore, orbit sub-Keplerian (and vice versa for an increasing pressure profile).
The change in gas density across the ring implies positive and negative radial pressure gradients at the inner and outer parts of the ring, respectively.
At the pressure maximum, $\eta$ becomes zero and the gas orbits with a Keplerian speed.
We note that the prescription for $u_\phi$ in Eq.~(\ref{equ:vphi}) only holds as long as the momentum feedback from dust grains is negligible \citep[a more complex solution when including dust feedback is given in][]{Benitez2019}.

\subsection{Gas drag and radiation pressure}
In protoplanetary disc studies, a standard way to characterise the dynamical behaviour of a particle is through the Stokes number, {\rm St}.
This dimensionless quantity compares the timescale over which friction forces the particle to move with the gas to the characteristic orbital timescale, $\Omega_{\rm K}^{-1}$ \citep[e.g.][]{Birnstiel2010}.
In the linear regime, which is applicable to both protoplanetary and debris discs, {\rm St} does not depend on the relative velocity between dust and gas.
It is commonly expressed as
\begin{equation}\label{equ:Stokes}
{\rm St} = \frac{\pi}{2}\,\frac{\rho_{\rm mat}\,a}{\Sigma_{\rm g}},
\end{equation}
where $\rho_{\rm mat}$ is the grain’s material density, $a$ is its physical radius, and $\Sigma_{\rm g}$ is the gas surface density.
The Stokes number thus quantifies the extent to which a dust particle is affected by gas drag, with the drag force often written as
\begin{equation}
{\bf f}_{\rm drag} = \frac{\Omega_{\rm K}}{\rm St}\left(\mathbf{u}-\mathbf{v}\right)\,,
\end{equation}
with the velocity vectors, ${\bf u}$ and ${\bf v}$ for gas and dust, respectively.
Eq.~(\ref{equ:Stokes}) shows that the coupling of a grain to the gas depends mainly on its size and the local abundance of gas.
Consequently, at very low gas densities, like those expected in the secondary gas scenario in debris discs, even small dust grains (${\sim}1\,\mu$m) can potentially only be weakly (or not at all) coupled to the gas flow.
To compute the momentum exchange between Lagrangian particles and gas that is modelled on a grid, two numerical steps are critical: first, the interpolation of the gas density and velocities from the Eulerian mesh to each particle’s position, and second the projection of the particle’s momentum feedback onto the gas, updating the gas velocities that are stored at the fixed interfaces between neighbouring grid cells.
We used the implemented cloud-in-cell method for interpolation.
In Appendix~\ref{appendix:feedback}, we compare the code's performance against analytical solutions for a radial drift test.
We further activated dust diffusion due to stochastic kicks between the dust particles and the gas \citep[][]{Fuente2017,Charnoz2011}, following \citet{Youdin2007} who derived an expression for the radial diffusivity of passive solid particles, $D_{\rm d }=\nu (1+4{\rm St}^2)(1+{\rm St}^2)^{-2}$.

In addition to frictional interactions with the gas and the central star’s gravitational force, we implemented the effect of radiation pressure on the dust dynamics.
We neglected the disc's self-gravity and the Poynting–Robertson drag \citep{Burns1979}, which operates on a much longer time scale than what is relevant in our simulations \citep[$t_{\rm P-R}{\sim}1\,\mathrm{Myr}$, cf. Eq.~(50) in][]{Burns1979}.
If the disc is optically thin to stellar irradiation, we can approximate the stellar flux at a certain location independently of the dust distribution.
Under this assumption, the radiation pressure force is given by
\begin{equation}
\mathbf{f}_{\rm rad} = \frac{\beta G M_\star}{r^2}\,\frac{\mathbf{r}}{\lvert \mathbf{r}\rvert}\,,
\end{equation}
where $\beta$ is a dimensionless parameter that depends on the grain’s material properties and the stellar parameters \citep{Burns1979}.
Most importantly, $\beta$ is a function of the particle's radius, with $\beta(a) = \beta_{1\mu\mathrm{m}}\times (1\mu{\rm m}/a)$, where $\beta_{1\mu{\rm m}}=\beta(a{=}1\mu{\rm m})$.

Notably, the radiation pressure force shares the same radial dependence as the star’s gravitational force,
\begin{equation}
\mathbf{f}_{\rm grav} = -\frac{GM_\star}{r^2}\,\frac{\mathbf{r}}{\lvert \mathbf{r}\rvert}\,.
\end{equation}
Because of this similarity, we can account for radiation pressure by replacing the original gravitational potential, $\phi_{\rm g}$, with an effective gravitational potential:
\begin{equation}\label{equ:red_grav_pot}
\phi_{\rm g}^\star = (1 - \beta)\,\phi_{\rm g}\,.
\end{equation}
In practice, we modified the code in a way that each dust particle experiences this effective gravitational potential, which depends on the particle size via $\beta(a)=\beta_{1\mu {\rm m}}\times (\mu{\rm m}/a)$.
For simplicity, we initialised the dust grains with their local Keplerian velocity.

\subsection{Numerical setup}
In the following section, we present results of one-dimensional (radial; Section \ref{subsec:1D}) and two-dimensional (radial/azimuthal; Section \ref{subsec:2D}) simulations.
The grid on which the gas density was evaluated extended from $0.1\,r_0$ to $3\,r_0$ with $N_r\!=\!512$ logarithmically spaced grid cells.
In the two-dimensional case, the azimuthal angle, $\varphi$, spanned from $-\pi$ to $\pi$ on $N_\varphi\!=\!1024$ cells.
We used wave killing zones \citep{devalBorro2006} at the inner and outer boundaries to prevent spurious reflections.

For the dust particles, to ensure equal sampling across different grain sizes, we drew superparticles from a logarithmically spaced size distribution using a uniform kernel.
Each superparticle was then assigned a representative mass \citep[similar to][]{Baruteau2019} assuming a dust number density profile of $n(a) \!\sim\! a^{-3.5}$ over the size range $a_{\min} \!=\! 0.8\,\mu\mathrm{m}$ to $a_{\max} \!=\! 1\,\mathrm{cm}$.
This distribution was motivated by that derived by \citet{Perrot2023} for \HD{} based on the SED fit.
For the material density of the particles, we used $\rho_{\rm mat}\!=\!3.3\,{\rm g\,cm}^{-3}$, which represents silicate grains.
For the initial setup, we positioned the particles inside the gas ring, between $0.9\,r_0$ and $1.1\,r_0$.
Our numerical treatment did not consider collisions between the dust particles themselves.
We ran the simulation for 200 orbits at $r_0$ and output density and velocity fields for every orbit.

\section{Results}\label{sec:results}
\subsection{One-dimensional case}\label{subsec:1D}
Dust trapping at gas-pressure maxima is well established in protoplanetary discs \citep[e.g.][]{Pinilla2012,Dullemond2018}. In contrast, debris discs are optically thin, leaving grains directly exposed to stellar radiation. Radiation pressure must therefore be included.
We began with 1D, axisymmetric simulations to examine how radiation pressure and gas drag set the radial structure of a coupled dust-gas ring before extending to more complex setups.

First, we note the differences to the scenario in which radiation pressure is not relevant, typical for protoplanetary discs.
As shown in Eq.~(\ref{equ:vphi}), the azimuthal gas velocity differs from the Keplerian velocity depending on the radial pressure gradient.
In the absence of radiation pressure, dust grains orbit with a Keplerian velocity.
This velocity mismatch between gas and dust generates friction, causing an exchange of angular momentum and leading to a radial drift of the dust grains towards regions of higher gas pressure \citep[e.g.][]{Whipple1972}.
As a result, in a ring-shaped gas distribution, dust grains accumulate at the maximum pressure, where both dust and gas rotate at the same (Keplerian) speed.

On the other hand, in optically thin debris discs, radiation pressure reduces the effective gravitational potential (Eq.~\ref{equ:red_grav_pot}), causing dust grains to orbit at
$v_{\varphi} \!=\! \sqrt{1 - \beta(a)}\,v_{\rm K}$.
Since $\beta(a) \!\geq\! 0$, the dust moves sub-Keplerian.
Additionally, if $\beta(a)\!>\! \eta$ the dust orbits slower than the gas, it experiences a headwind, and this frictional interaction transfers angular momentum from the gas to the dust.
As a result, the dust drifts outward to regions of lower gas pressure.
This drift continues until $\beta(a) \!=\! \eta$, at which point no further angular momentum exchange occurs \citep[see][for a detailed discussion]{Takeuchi2001}.
To conserve angular momentum, the gas moves in the opposite direction (i.e. inward where dust drifts outward).

\begin{figure*} 
    \centering
    \includegraphics[width=\textwidth]{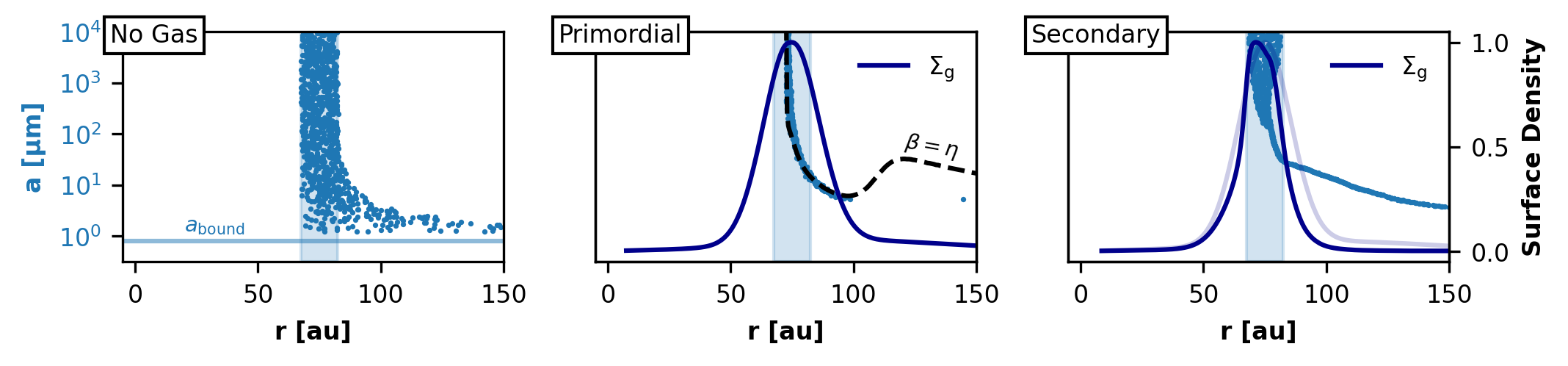}
    \caption{One-dimensional simulations of a dust ring. All setups included radiation pressure with $\beta_{1\mu{\rm m}} \!=\! 0.4$, and are shown after 150 orbits at $r_0$. {\it Left:} Simulation without gas. The horizontal line indicates the minimum particle size for which the dust remains on bound orbits, and the shaded area marks the initial radial range of the dust particles. {\it Centre:} Includes gas density specified by Eq.~(\ref{equ:dens0}), with gas density ($\Sigma_\mathrm{g}$, blue curve) normalised to its maximum value, which was set to represent a gas ring of $10\,M_\oplus$, representing a primordial scenario. The dust-to-gas ratio is 0.01. The dashed curve shows where $\beta \!=\! \eta$. {\it Right:} Same as the centre panel, but with $M_{\rm gas}\!=\!0.1\,M_\oplus$, and a dust-to-gas ratio of unity, reflecting a scenario of secondary gas origin. Large particles are affected by neither radiation pressure nor gas drag, and they remain on their initial circular Keplerian orbits. For sizes $\lesssim\!100\,\mu\mathrm{m}$, both gas drag and radiation pressure become efficient, and the particles move outward. The momentum feedback from the dust changes the gas density profile from its initial distribution (transparent blue line).
    }
    \label{fig:1D}
\end{figure*}
To test these expectations, we performed one-dimensional simulations of dust co-located with a gas ring using Dusty FARGO-ADSG with just one azimuthal cell.
We represented the dust using 1,000 particles, initialised around the peak of the gas density distribution, and ascribed a Keplerian velocity.
The ring is centred at $r_0 \!=\! 75\,\mathrm{au}$, matching the location observed in \HD{} \citep[see][]{overview_arks,gas_arks}.
We set $\beta_{1\mu\mathrm{m}} \!=\! 0.4$, which corresponds to a blow-out size (particles for which $\beta\!=\!0.5$) of $0.8\,\mu\mathrm{m}$, approximately the observationally inferred minimum grain size \citep[][see also Sec.~\ref{subsec:1D}]{Perrot2023}.
We then considered three scenarios: (i) a gas-free disc, where particles were influenced solely by gravity and radiation pressure, (ii) a gas-rich disc with a total gas mass of $10\,M_\oplus$, a molecular weight of $\mu\!=\!2.3$, and a dust-to-gas ratio of 0.01, representative of the primordial scenario, where particles felt the gas drag but momentum feedback from dust to gas was small, and (iii) a gas-reduced disc with a total gas mass of $0.1\,M_\oplus$, a molecular weight of $\mu\!=\!14$, and a dust-to-gas ratio of unity, representative of a secondary gas scenario.
When included, the gas ring was characterised by a temperature of $T(r_0)\!=\!38\,{\rm K}$ \citep[motivated by][]{line_arks}, a radial width of $\sigma \!=\! 0.14\times r_0$,  a contrast to the background of $\mathcal{C} \!=\! 20$, and a viscosity level of $\alpha\!=\!10^{-4}$.
As both gaseous scenarios used the same temperature at different molecular weights, the aspect ratio was different in the two cases.
Scenario (ii) was characterised by an aspect ratio of $h(r_0)\!=\!0.07$, while scenario (iii) used $h(r_0)\!=\!0.028$.
This setup produced a maximum surface density of $\Sigma_{\mathrm{max}} \!=\! 0.17\,\mathrm{g\,cm}^{-2}$ for case (ii) and $\Sigma_{\mathrm{max}} \!=\! 3.1 \times 10^{-3}\,\mathrm{g\,cm}^{-2}$ for case (iii).
Each simulation was evaluated after 150\,orbits at $r_0$, corresponding to approximately 71\,kyr.

Fig.~\ref{fig:1D} shows the particles' radial distance from the star, with the vertical axis indicating the particle size.
The left panel shows the gas-free scenario.
Because radiation pressure depends on particle size, only the smallest grains ($a \!\lesssim\! 10\,\mu{\rm m}$) deviate significantly from their initial radial region (shaded area).
The horizontal line marks the blowout size $a_{\rm bound} \!=\! 0.8\,\mu\mathrm{m}$ (at which $\beta(a) \!=\! 0.5$), above which particles remain on bound orbits.
Particles smaller than $a_{\rm bound}$ escape the system, while those just above $a_{\rm bound}$ follow highly eccentric orbits.
The resulting broad spread in the radial distance for particles of $a_{\rm bound} \!\leq\! a \!\lesssim\! 10\,\mu{\rm m}$ arises from these high eccentricities, with the simulation snapshot capturing each particle at a different point in its eccentric orbit.

The central panel of Fig.~\ref{fig:1D} corresponds to case (ii), where we examined a gas ring that encases the dust grains, where the dust-to-gas mass ratio is $f_{\rm dtg}\!=\!0.01$.
The shaded region indicates the initial radial extent of the dust particles, while the solid blue curve shows the gas surface density after 150\,orbits.
For each radial cell, the particle size at which $\beta(a) \!=\! \eta(r)$ is indicated by the black dashed line.
This simulation reveals three distinct regimes of particle behaviour.
First, the largest particles ($\gtrsim\!100\,\mu{\rm m}$) remain mostly unaffected by radiation pressure. 
Beyond the pressure maximum, the gas orbits sub-Keplerian because the inward pressure gradient implies $\eta \!>\! 0$ (see Eq.~\ref{equ:vphi}).
The large particles, for which $\beta\!<\!\eta$, orbit faster than the gas, thus, they lose angular momentum due to friction. 
Consequently, they drift towards the pressure maximum at the ring's radial centre.

Particles of $\lesssim\! 50\,\mu\mathrm{m}$, start to be affected by radiation pressure and thus orbit with a sub-Keplerian velocity.
For those particles, there is a location where $\beta(a)\!>\!\eta$ and the particles orbit slower than the gas and thus experience a tailwind.
This induces an outward drift until they reach a location where $\beta\!=\!\eta$ (marked by the dashed line).
Notably, dust grains do not maintain significant eccentricities here because gas drag efficiently damps any non-circular motion.

For particles of $\lesssim\!5\,\mu\mathrm{m}$ the relation $\beta\!>\!\eta$ holds everywhere in the domain; these grains experience a net outward drift with no equilibrium point and thus leave the domain through the outer boundary.
In this primordial–gas scenario, the minimum size that remains bound in the ring is therefore substantially larger than in the gas–free case (where $a_{\rm bound}\!\simeq\!0.8\,\mu\mathrm{m}$), because gas drag transfers angular momentum to the grains and assists their outward motion.
The grains that do remain trapped align with the local balance condition $\beta(a)\!=\!\eta$.
As $\eta$ is size-independent, changing the radiation pressure normalisation $\beta_{1\mu{\rm m}}$ by a certain factor simply shifts the dashed line in the central panel vertically by the same factor.
This yields three immediate implications:
(i) the minimum retained size is set by the threshold $\beta(a_{\min})\!=\!\eta$; for compact grains with $\beta\!\sim\! a^{-1}$, this implies $a_{\min}\!\sim\! \eta^{-1}$;
(ii) a warmer ring (larger sound speed $c_{\rm s}$ and hence larger aspect ratio $h$) retains smaller grains, because $\eta\!\sim\! h^{2}$;
(iii) a narrower ring (steeper radial pressure gradient) also retains smaller grains, since $\eta \!\sim\! h^{2}\,|\mathrm{d}\ln P/\mathrm{d}\ln r|$ increases with the pressure slope.

Finally, the right panel of Fig.~\ref{fig:1D} illustrates the outcome for scenario (iii), a secondary origin gas-poor disc, with a dust-to-gas ratio of unity.
Here, particles $\gtrsim\!1\,{\rm cm}$ are neither affected by radiation pressure nor gas drag, on the simulated time scales, and remain on their initial circular Keplerian orbits.
For sizes $\lesssim30\,\mu\mathrm{m}$, both the gas drag and radiation pressure become efficient and the particles move outward.
The momentum feedback from the dust changes the gas density profile from its initial distribution (transparent blue line).
As the dust orbits even slower than the gas, it receives angular momentum from the gas via friction.
Hence, as the gas loses angular momentum, it is forced to move inwards, steepening the outer slope of the density distribution.
We highlight that once dust feedback becomes significant, the azimuthal gas velocity is modified by the friction and Eq.~(\ref{equ:vphi}) does no longer apply.
Hence, the equilibrium location is no longer represented by the $\beta\!=\!\eta$-line.

\subsection{Two-dimensional fiducial model}\label{subsec:2D}
In this section, we investigate how the system develops under an azimuthally asymmetric instability.
To this end, we need a mechanism that triggers an azimuthally asymmetric profile in the gas structure.
Here, we used as an initial condition an idealised Gaussian-like gas density profile (described in Eq.~\ref{equ:dens0}) that is tailored to trigger such instability, when the azimuthal coordinate is included in the simulation.
In particular, we set the parameters of the simulation such that the ring is susceptible to the Rossby Wave Instability \citep[RWI,][]{Lovelace1999,Ono2016,Chang2023}.
The occurrence of this instability can be controlled by the width, $\sigma$, of the density profile in Eq.~(\ref{equ:dens0}).
\citet{Ono2018} presented a dedicated study that analyses the conditions and evolution of the RWI in a gas ring and which azimuthal mode is most unstable.
We emphasise that the results by \citet{Ono2018} are relevant but not directly applicable, as we include the momentum feedback from the dust onto the gas in our simulation.
This was not included in their study;
it changes the equations of motion and adds a dependency on the dust-to-gas ratio to the gas dynamics.
We adopted a viscosity level of $\alpha=10^{-4}$.
The (magneto)hydrodynamical processes that transport angular momentum in protoplanetary discs \citep{Lesur2023,Rosotti2023} may partially apply to debris-disc gas as well \citep{Cui2024}.
In the low-density regime, \citet{Cui2024} further argue that molecular viscosity can become non-negligible.
For our fiducial model this is probably not relevant as we estimate a molecular viscosity level of $\alpha\sim 10^{-7}$, but this contribution increases linearly with decreasing gas mass.
The magnetorotational instability \citep[MRI,][]{Balbus1991} is not expected to be active in the midplane at the densities of our fiducial model \citep{Cui2024}; but turbulence generated in upper layers could be transported there.
Overall $\alpha = 10^{-4}$ should be regarded as a conservative upper limit for the scenarios considered here.
We note, however, that at significantly lower gas densities than our fiducial case, $\alpha$ could be significantly larger because both the molecular viscosity and MRI-turbulence contribution increase.

In the linear regime, the RWI grows rapidly, typically within $10\!-\!20$ dynamical periods, and is only weakly affected by viscosity \citep{Lin2014}. Once the non-linear regime is reached, however, viscosity plays a central role in setting the vortex lifetime, with higher $\alpha$ accelerating vortex dissipation \citep{Lin2014}.

As the dust mass can be estimated from the spectral energy distribution \citep[$0.21\,M_\oplus$;][]{overview_arks,Perrot2023}, and the gas temperature from the CO molecular line emission \citep[38\,K at $r_0$;][]{line_arks}, we keep these values fixed.
We model the dust population using $10^5$ superparticles.
When considering a molecular mass of $\mu\!=\!2.3$ this temperature corresponds to an aspect ratio of $h_0\!=\!0.07$.
We found that the ring becomes unstable when its radial width is $\sigma\!\lesssim\!1.5\!\times\! H_{\rm p}$, with $H_{\rm p}\!=\!h\times r$ being the gas pressure scale height.
In the fiducial model we therefore set a moderately unstable width of $\sigma\!=\!1.4\!\times\! H_{\rm p}(r_0) \!=\! 1.4 \!\times\! h_0 r_0 \!=\! 7.4\,{\rm au}$.
This is very similar to the condition found in \citet{line_arks} that the FWHM of the radial gas density distribution must be narrower than $\sim\!17.3\,$au, corresponding to a width of $\sigma\!=\!7.3\,$au.
We defined the ratio between the estimated CO mass \citep[$m_{\rm CO}\!=\!0.15\,M_\oplus$,][]{line_arks} and the total gas mass, $\chi_{\rm CO} \!\equiv\! m_{\rm CO}/m_{\rm gas}$.
The typical value adopted for protoplanetary discs, $\chi_{\rm CO} \!\sim\! 10^{-3}$ \citep[e.g.][]{Williams2014}, is inherited from the interstellar medium.
However, the fractional abundance of CO is highly uncertain and depends on thermal and chemical processes during the disc's lifetime.
Direct measurements of hydrogen have found that CO tends to be reduced in evolved discs \citep[e.g.][]{Bergin2013,McClure2016,Trapman2017}, indicating depletion caused by freeze-out, photodissociation, and chemical reactions.  
Finally, the CO mass fraction also depends on whether the gas is of primordial or secondary origin -- or a mixture.
Table~\ref{tab:fiducial} lists the parameters that we use for the fiducial run.
Note that some of these parameters are interdependent, such as the dust-to-gas ratio and the CO mass ratio, or the temperature and the aspect ratio.
\begin{table}
    \caption{Fiducial model parameters adopted in Sec.~\ref{subsec:2D}.}
    \label{tab:fiducial}
    \centering
    \begin{tabular}{l c c}
    \hline\hline
    Quantity & Parameter & Value \\
    \hline
    Viscosity              & $\alpha$          & $10^{-4}$     \\
    Ring Gas Density & $\Sigma_{0}$ &  $0.12 \,{\rm g}\,{\rm cm}^{-2}$             \\
    Background Density & $\Sigma_{\rm bg}$ &  $6\times 10^{-3} \,{\rm g}\,{\rm cm}^{-2}$             \\
    Density Contrast   & $\mathcal{C}$     & 20 \\
    Ring location& $r_0$             & 75\,au \\
    Ring width              & $\sigma$     & 7.4\,au \\
    Background peak location & $r_{\rm bg}$             & 103\,au \\
    Background width              & $\sigma_{\rm bg}$     & 43\,au \\
    Total gas mass & $M_{\rm gas}$ & $50\,M_\oplus$\\
    Temperature at $r_0$ & $T_0$ & 38\,K\\
        Aspect Ratio           & $h$                 & 0.07 \\
    Dust material density  & $\rho_{\rm mat}$  & 3.3\,g\,cm$^{-3}$ \\
    Radiation strength     & $\beta_{1\mu{\rm m}}$           & 0.4 \\
    Dust-to-gas ratio       &   $f_{\rm dtg}$     & $4.2\times10^{-3}$ \\
    Mean-molecular weight & $\mu$ & 2.3 \\
    CO mass ratio & $\chi_{\rm CO}$ & $3\times10^{-3}$\\
    \hline
    \end{tabular}
\end{table}

\begin{figure*}
    \sidecaption
    \includegraphics[width=12cm]{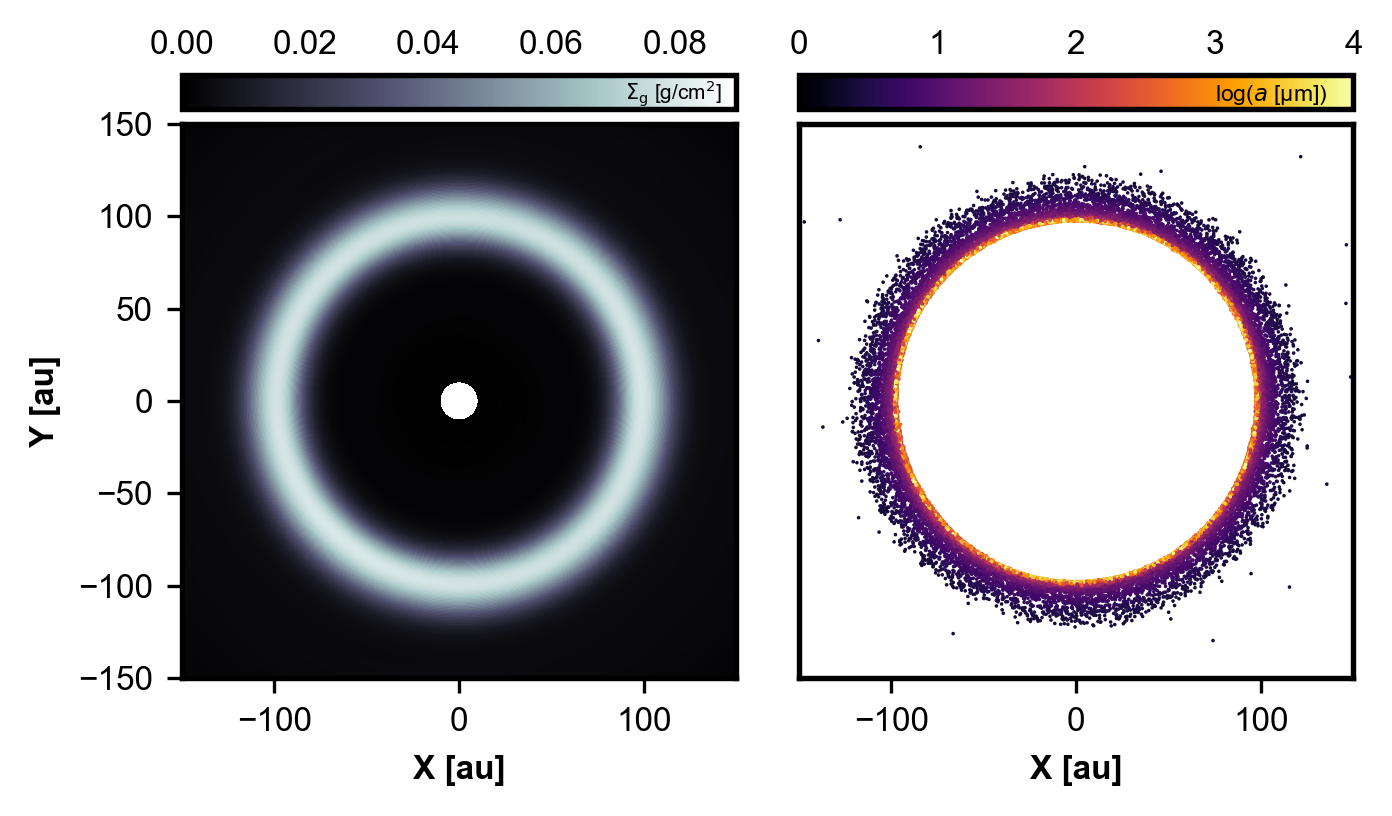}
    \caption{Fiducial hydrodynamical model after 150 orbits at $r_0$ (corresponding to about 71\,kyr). The total gas mass of the Gaussian ring is $M_{\rm g} \!=\! 50\,M_\oplus$, with a dust-to-gas mass ratio of $4.2\!\times\!10^{-3}$. {\it Left:} Gas surface density distribution. {\it Right:} Particle distribution with colour indicating the grain size.}
    \label{fig:hydro_fiducial}
\end{figure*}

\begin{figure}
    \centering
    \includegraphics[width=\columnwidth]{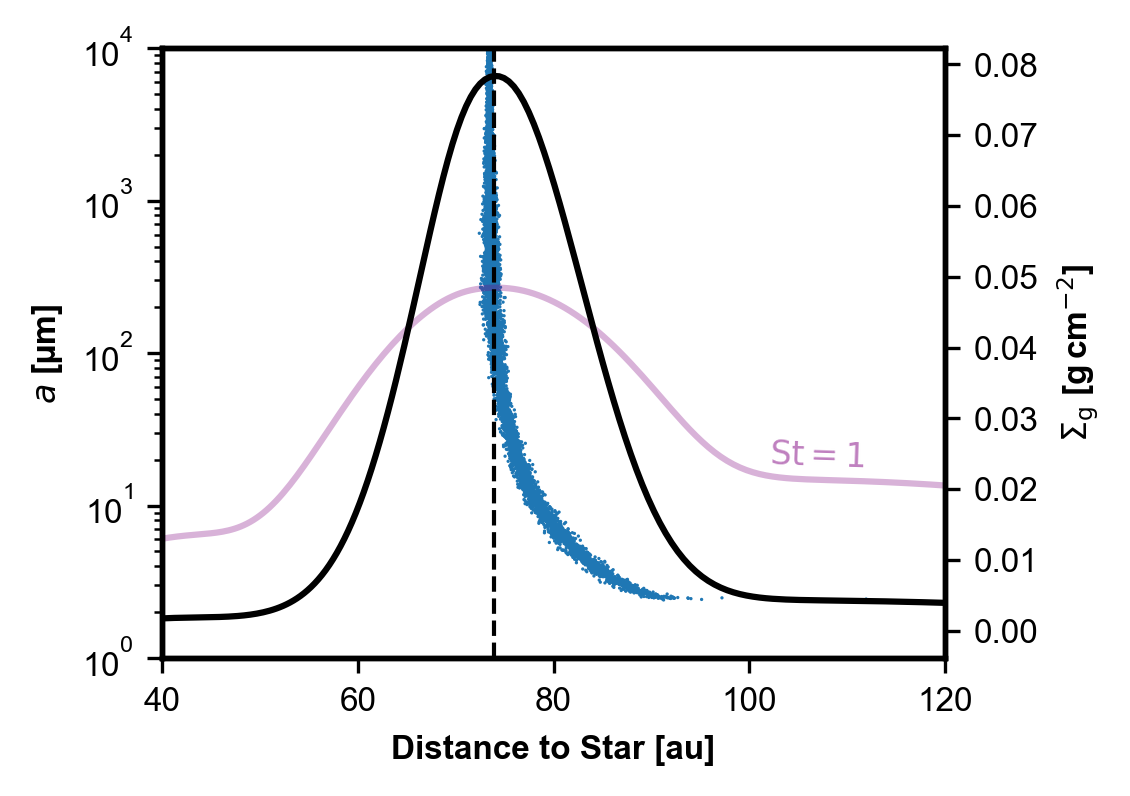}
    \caption{Radial distribution of particles corresponding to Fig.~\ref{fig:hydro_fiducial}. The figure incorporates every tenth particle. The black curve traces the azimuthally-averaged gas surface density profile. The vertical dashed line marks the pressure maximum.
    The purple curve indicates the particle size that corresponds to a Stokes number of unity.}
    \label{fig:rad_profile}
\end{figure}

\begin{figure}
    \centering
    \includegraphics[width=\columnwidth]{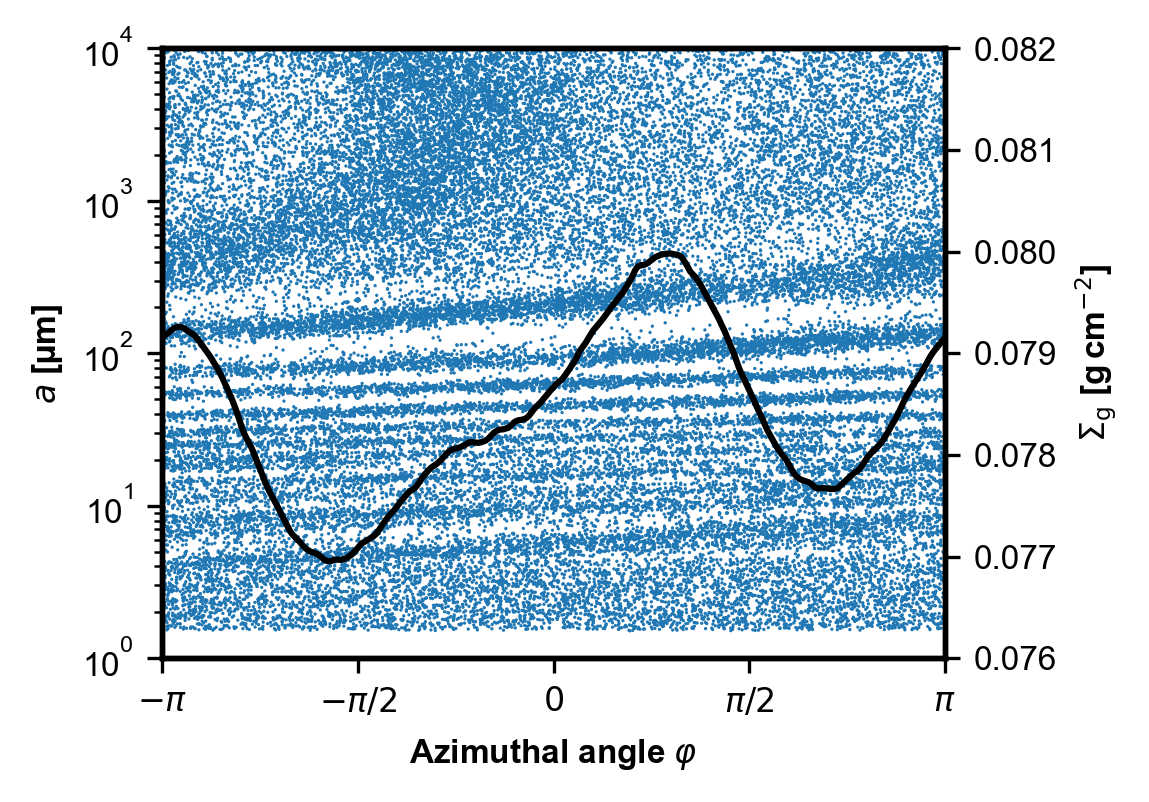}
    \caption{Azimuthal distribution of particles of different sizes corresponding to Fig.~\ref{fig:hydro_fiducial}. The figure incorporates every tenth particle. The black curve traces the radial peak value of the gas surface density and corresponds to the right y-axis.}
    \label{fig:azi_profile}
\end{figure}

In the fiducial simulation, a vortex emerges after around 10 orbits at $r_0$ ${\sim}4.7\,{\rm kyr}$.
The instability grows rapidly until 25 orbits (${\sim}11.8\,{\rm kyr}$), but then starts to abate.
The system reaches an azimuthally symmetric state after about 200 orbits (${\sim}94\,{\rm kyr}$).
We show a time series of the azimuthal asymmetry caused by the instability in Appendix~\ref{appendix:timeseries}.

The left panel of Fig.~\ref{fig:hydro_fiducial} shows the gas surface density distribution after 150\,orbits at $r_0$ (${\sim}71\,{\rm kyr}$), while the right panel shows the particle positions with the colour indicating the grain size.
The size-dependent distribution of the solids is better illustrated by the radial and azimuthal profiles in Figs.~\ref{fig:rad_profile} and \ref{fig:azi_profile}, respectively.

Figure~\ref{fig:rad_profile} shows that dust grains of about $30\,\mu\mathrm{m}$ and larger are efficiently trapped at the pressure maximum of the gas ring, indicated by the vertical dashed line.
The purple curve in Fig.~\ref{fig:rad_profile} indicates where $\mathrm{St}\!=\!1$ (roughly $250\,\mu\mathrm{m}$ at the peak).
For particles with \(a \lesssim 30\,\mu\mathrm{m}\), radiation pressure becomes increasingly significant, shifting the distribution slightly outward relative to larger grains.

Fig.~\ref{fig:azi_profile}  highlights the distinct behaviours of the dust grains more clearly. Namely, the azimuthal distribution reveals three regimes: (i) small grains (\(a \!\lesssim\! 10\,\mu\mathrm{m}\)) that closely trace the azimuthal gas structure, (ii) partially decoupled grains (\(10\,\mu\mathrm{m} \!\lesssim\! a \!\lesssim\! 300\,\mu\mathrm{m}\)) that are still strongly affected by the gas flow, and (iii) very decoupled grains (\(a \!\gtrsim\! 300\,\mu\mathrm{m}\)) that are affected by the gas flow only on extended timescales.
Between $10\,\mu\mathrm{m}$ and $300\,\mu\mathrm{m}$, the response to the azimuthal gas variations is noticeably size-dependent.
That dust particles of different Stokes number are trapped at different azimuths in the case of a strong vortex has been investigated in detail in several studies \citep{Baruteau2016,Fuente2017,Hammer2019}. 
The shift away from the gas maximum in larger particles is attributable to their increased inertia relative to the gas flow, which corresponds to a longer frictional stopping time.

\begin{figure}
    \centering
    \includegraphics[width=\columnwidth]{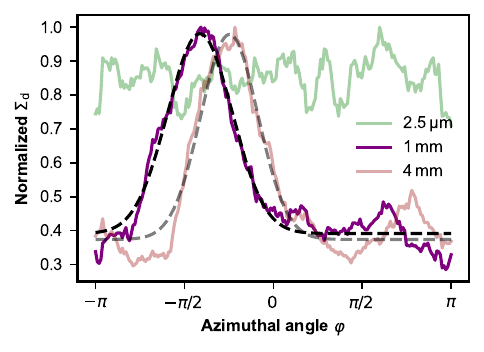}
    \caption{Azimuthal dust distribution for representative sizes of 2.5$\,\mu$m, 1$\,$mm and 4\,mm. The profiles were produced by selecting the particles within $\pm10\%$ of the respective size over the full radial domain and normalising to the maximum value of the large grains. The dashed black curve represents a Gaussian fit to the data 1\,mm profile.}
    \label{fig:azi_profile2}
\end{figure}

To study the azimuthal distribution more quantitatively, we present three representative normalised dust density profiles in Fig.~\ref{fig:azi_profile2}.
Here, we binned the particles around $2.5\,\mu\mathrm{m}$ (${\rm St}\sim0.01$), 1\,mm (${\rm St}\sim3.5$) and $4\,\mathrm{mm}$ (${\rm St}\sim14$), respectively, summing those within $\pm 10\%$ of each central particle size.
The figure shows that, apart from noise, the small particles are azimuthally spread.
The two larger species are azimuthally confined but the 4\,mm particles are shifted in the positive azimuthal direction with respect to the 1\,mm grains.
Gaussian fits to the 1\,mm and 4\,mm distributions are plotted as dashed black curves in Fig.~\ref{fig:azi_profile2}. 
The fits yield widths of $\sigma_{1\mathrm{mm}} \!=\! 32.0^\circ \pm 0.2^\circ$ and $\sigma_{4\mathrm{mm}} \!=\! 28.9^\circ \!\pm\! 0.2^\circ$, and an azimuthal shift between the two species of $30.0^\circ \!\pm\! 0.3^\circ$.
This is a much larger shift than that found in \citet{Baruteau2016} for the corresponding Stokes numbers and is due to the fact that at the time plotted the gas vortex has weaken due to dissipative effects (see Appendix~\ref{appendix:timeseries}).
Similar to the azimuthal asymmetry in the gas, the azimuthal accumulation of dust grains is also time-dependent and both the shift and the contrast of the asymmetry depend on the snapshot at which the simulation is evaluated.

Grain size transitions in gas coupling depend on the gas surface density, and thus on the total gas mass. Effects of varying the gas mass (while all else is kept the same, including the dust mass) is discussed next in Section~\ref{sec:RT}.

\section{Radiative transfer and image synthesis}\label{sec:RT}
We employed the radiative transfer code RADMC-3D \citep[version 2.0;][]{Dullemond2012}, a grid-based method that requires density fields as input. 
Accordingly, we constructed a three-dimensional spherical grid with $N_{r}\times N_{\theta}\times N_{\varphi} = 512 \times 30 \times 1024$ cells.
For the dust component, we separated the superparticles into fifty logarithmically spaced representative size bins and transformed the discrete superparticle distribution into a grid-based dust density via cloud-in-cell interpolation.
Because our hydrodynamical simulations are two-dimensional, we first computed the dust surface density $\Sigma_{\mathrm{d}}(r,\varphi)$ and then extended the distribution in the vertical direction according to the following:
\begin{equation}
\rho_{\mathrm{d}}(r,\varphi,z) \;=\; \frac{\Sigma_{\mathrm{d}}(r,\varphi)}{\sqrt{2\pi}\,H_{\mathrm{d}}}\;\exp\!\left(-\tfrac{z^{2}}{2\,H_{\mathrm{d}}^{2}}\right)\,,
\end{equation}
where the dust scale height, following \citet{Dubrulle1995}, is given by
\begin{equation}
H_{\mathrm{d}} \;=\; \sqrt{\frac{\alpha_{\mathrm{D}}}{\alpha_{\mathrm{D}} + \mathrm{St}}}\;H_{\mathrm{g}}\,.
\end{equation}
We assumed that the diffusion coefficient $\alpha_{\rm D}$ is equal to the viscous $\alpha$ parameter used in the simulations, which in our case served as a numerical convenience rather than a physically constrained quantity.
While this approach does not yield physically motivated dust scale heights, we found that the choice has little impact on the resulting synthetic images, as most of the emission arises near the disc midplane where the vertical dust distribution is densest.
We computed the opacities for each representative dust size using the {\tt optool}\footnote{\url{https://github.com/cdominik/optool}} package \citep{Dominik2021},
assuming compact grains with a material composition of pyroxene silicates.
We then used RADMC-3D to compute the dust temperature with the {\tt mctherm} task, employing $n_{\rm phot} \!=\! 10^8$ photon packages.
Because the disc is optically thin, the dust temperature varies only weakly with height: grains at virtually all altitudes are directly exposed to stellar radiation.
At the ring radius, grains larger than 10$\,\mu{\rm m}$ are of ${\sim}\,62\,{\rm K}$, essentially the black-body equilibrium temperature reported by \citet{gas_arks}.
Smaller grains absorb stellar photons more efficiently than they re-emit at long wavelengths and therefore become slightly hotter; the smallest grains we consider ($0.8\,\mu{\rm m}$) reach roughly 65\,K at the same location.
Next, we performed ray-tracing simulations of the thermal emission at $\lambda_{\rm obs} \!=\! 0.89\,\mathrm{mm}$ and of scattered stellar light at $\lambda_{\rm obs} \!=\! 1.25\,\mu\mathrm{m}$, using anisotropic scattering based on the scattering matrices calculated with {\tt optool}.
We assumed an inclination of 44.1$^\circ$ and a position angle of $58.7^\circ$, as reported in \citet{overview_arks}.

To generate more realistic comparisons with actual observations, we applied image synthesis techniques to replicate each telescope’s finite resolution and sensitivity.
In the case of ALMA band~7, we used the CASA {\tt simobserve} task, including realistic noise contributions from atmospheric fluctuations and antenna responses.
We found a good match with the observational beam when using a combination of configurations C-5 (with integration time $\tau_{\rm int}\!=\!345\,{\rm min}$) and C-3 ($\tau_{\rm int}\!=\!84\,{\rm min}$) and weights of ${\tt robust\!=\!0.5}$.

For the polarised $J$-band image, we convolved the model output with the diffraction-limited resolution of an 8.2\,m telescope at $\lambda_{\mathrm{obs}}\!=\! 1.25\,\mu\mathrm{m}$, then added Poisson noise to approximately match the observed signal-to-noise ratio.

\begin{figure*}
    \sidecaption
    \includegraphics[width=12cm]{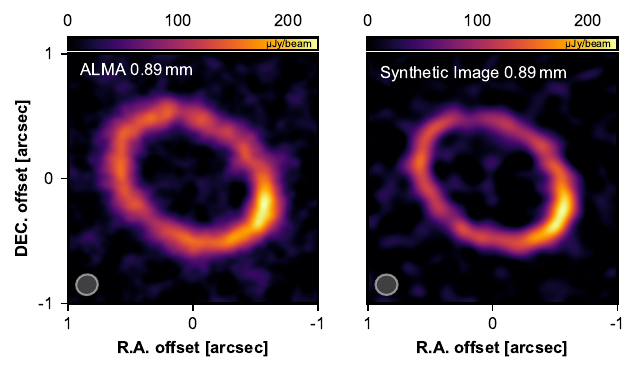}
    \caption{Comparison between ALMA observation and simulation. {\it Left:} ALMA band\,7 observation of \HD{} \citep[][]{overview_arks,hd121617_arks}. {\it Right:} Synthetic image of the fiducial hydrodynamical model at 0.89\,mm using {\tt CASA}-simobserve. In both images, the beam is shown as a grey ellipse in the bottom left corner.}
    \label{fig:simobserve}
\end{figure*}

\begin{figure*}
    \sidecaption
    \includegraphics[width=12cm]{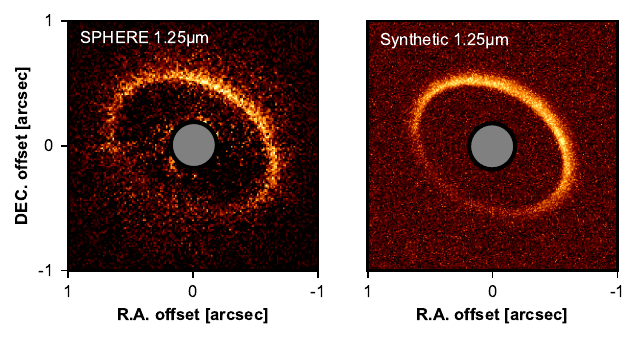}
\caption{Comparison between scattered light observation and simulation. {\it Left:} VLT/SPHERE $J$-band ($\lambda_{\rm obs}\!=\!1.25\,\mu{\rm m}$) observation of \HD{} \citep[see][]{scat_arks}. {\it Right:} Synthetic image of polarised light at $1.25\,\mu{\rm m}$ for the fiducial hydrodynamical model. In both panels, the grey circle indicates the coronagraphic mask of the observation.}
    \label{fig:Jband}
\end{figure*}

\subsection{Fiducial model}

Fig.~\ref{fig:simobserve} compares the ALMA band~7 observation of \HD{} \citep{overview_arks} with the corresponding synthetic image of our hydrodynamical model.
Qualitatively, the simulated image represents the observation to a high degree; in particular, the ring's radial extent and the arc's azimuthal distribution and contrast are well reproduced.
In a more detailed inspection, the right panel reveals a brighter arc in the ring where dust grains are azimuthally concentrated, superimposed on an azimuthally symmetric emission component.
This combination results in a slightly stronger contrast between the arc and the rest of the ring (see Sec.~\ref{subsec:observe_predictions}) in the synthetic image compared to the real observation.

Fig.~\ref{fig:Jband} presents the VLT/SPHERE polarised $J$-band image of \HD{} \citep{scat_arks} alongside its simulated counterpart.
Polarised $J$-band observations primarily capture starlight scattered off small dust grains in the circumstellar environment.
Since small particles (\(a \!\lesssim\! 10\,\mu\mathrm{m}\)) follow the largely azimuthally symmetric gas distribution, the brightness pattern is dominated by angle-dependent scattering efficiency and phase function, rather than azimuthal density variations.
Even in polarised light, forward scattering is typically enhanced, while backward scattering is suppressed \citep[e.g.][]{Min2012}.
As a result, the near (northwest) side of the ring, where forward scattering prevails, appears brighter, while the far side is scarcely visible.
We note that the simulated scattered light image appears more radially extended than the actual observation, which may indicate an overestimate of dust diffusion in our model. 

As previously stated, in addition to the azimuthal trapping of mm grains, the most important effect of the gas drag is the radial mismatch of the ring observed by ALMA ($r=73.6\,{\rm au}$) and in scattered light ($r=80.6\,{\rm au}$) \citep{scat_arks}. This shift corresponds to 9.5\%.
\begin{figure}
    \centering
    \includegraphics[width=\columnwidth]{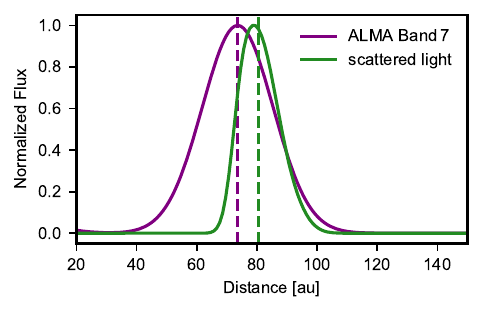}
    \caption{Deprojected radial intensity profiles of ALMA band\,7 synthetic observation (purple, corresponding to right panel of Fig.~\ref{fig:simobserve}) and $J$-band scattered light (green, corresponding to right panel of Fig.~\ref{fig:Jband}). The vertical dashed lines mark the corresponding maxima measured from the real observations at $73.6\,{\rm au}$ and $80.6\,{\rm au}$ for sub-mm emission and scattered light, respectively \citep[see][]{scat_arks}.}
    \label{fig:rad_prof_scat_mm}
\end{figure}
Fig.~\ref{fig:rad_prof_scat_mm} shows the deprojected radial intensity profiles of the synthetic ALMA band\,7 observation (cf. right panel of Fig.~\ref{fig:simobserve}) and the synthetic scattered light image (cf. right panel of Fig.~\ref{fig:Jband}), both normalised to their respective maximum values.
The dashed lines in Fig.~\ref{fig:rad_prof_scat_mm} mark the measured ring locations from the observations \citep[][]{scat_arks}.
By fitting Gaussians to the radial profiles of the synthetic images, we find the scattered light ring to be shifted outward by 8.2\% with respect to the ALMA ring. This is qualitatively in good agreement with the real observations 

\subsection{The gas mass upper limit}
We attribute the offset between ALMA and SPHERE rings to radiation pressure, which predominantly affects small dust grains.
This additional force changes their radial force balance, resulting in a sub-Keplerian equilibrium orbital velocity.
However, for these small grains to be radially displaced from the gas, they must be at least partially decoupled from the gas dynamics; otherwise, any relative motion with respect to the gas dynamics will be immediately dampened.
We demonstrate this with a simulation in which the gas mass is increased by a factor of five relative to our fiducial model (from $50\,M_\oplus$ to $250\,M_\oplus$): although the ALMA arc can still be reproduced at this higher gas mass, particles of $a\lesssim 10\,\mu$m are tightly coupled to the gas dynamics.

\begin{figure}
    \centering
    \includegraphics[width=\columnwidth]{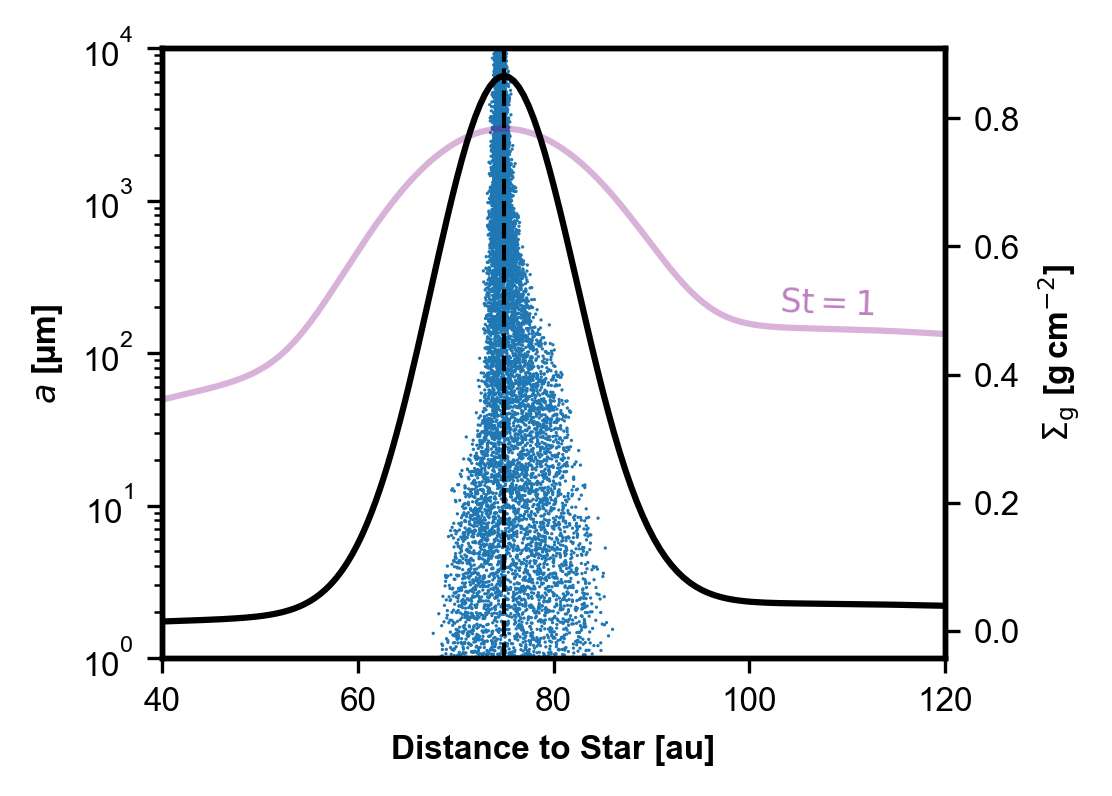}
    \caption{Radial particle distribution for different sizes from a two-dimensional hydrodynamical simulation with $M_{\rm gas}\!=\!250M_\oplus$.
    The vertical dashed line marks the ring location at 75\,au. The purple profile marks the particle sizes that correspond to ${\rm St}\!=\!1$.}
    \label{fig:uppermass}
\end{figure}
Fig.~\ref{fig:uppermass} shows the radial distribution of the particles.
Compared to the distribution of the fiducial model shown in Fig.~\ref{fig:rad_profile}, particles of $a\!\lesssim\!10\,\mu$m are much more radially spread in the case of increased gas mass.
Additionally, those particles extend to inside the pressure maximum (marked by the vertical dashed line).
The radiation pressure does not dislocate the particles as significantly as in the fiducial model.
Gaussian fits to the rings indicate that the peak of the scattered light ring lies only 0.8\% farther out than that of the ALMA band\,7 ring.
\begin{figure}
    \centering
    \includegraphics[width=0.8\columnwidth]{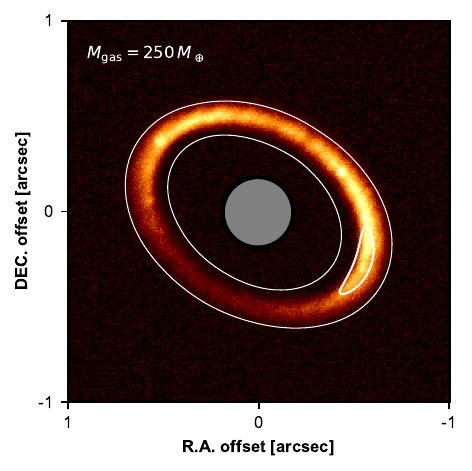}
    \caption{Same as right panel of Fig.~\ref{fig:Jband} but for the model that assumes a higher gas mass, with $M_{\rm gas}\!=\!250\,M_\oplus$.
    The contours delineate the ALMA band\,7 continuum emission.
    The model fails to reproduce the characteristic features of the $J$-band image (cf. right panel of Fig.~\ref{fig:obs}), i.e. the radial dislocation with respect to the ALMA ring, and the ring's sharpness.}
    \label{fig:Jband_upper}
\end{figure}
This model of increased gas density would fail to reproduce both the outward shift and the sharpness of the ring seen in scattered light.
We demonstrate this by producing the $J$-band polarised light image for the increased gas mass, shown in  Fig~\ref{fig:Jband_upper}.
The image shows the corresponding ALMA band\,7 continuum emission overlaid as white contours.
This shows a rough agreement between the scattered light and sub-mm ring radii, inconsistent with the observations.
Additionally, it shows that the scattered light ring is more radially spread.

Therefore, the upper limit for the gas mass is between the value used in the fiducial model (50\,$M_{\oplus}$) and the 250\,$M_\oplus$ analysed here.
In particular, this demonstrates that the ring around \HD{} is not consistent with the gas content of a typical Class II disc around a 1.9\,$M_\odot$ star \citep[$\!\sim\!1\,M_{\rm J}$, e.g.][]{Ansdell2016}.

\subsection{The gas mass lower limit}\label{subsec:lowlimit}
\subsubsection{$M_{\rm gas}\!=\!M_{\rm CO}$}
First, we probed the scenario where the reported CO mass of 0.15\,$M_\oplus$ represents the total gas mass, $M_{\rm gas}\!=\!0.15\,M_\oplus$.
This change affects the ring dynamics in two ways.
Every grain’s Stokes number rises by more than two orders of magnitude compared to the fiducial model.
Particles that were mildly influenced by drag in the fiducial model now move independently, while previously well-coupled grains become only weakly coupled.
Second, with the total dust mass held constant, the dust-to-gas ratio in the ring increases to above unity, $f_{\rm dtg}\!=\!1.4$.
\citet{Crnkovic2015} found that once dust accumulates in a vortex the dust feedback destroys the vortex if the trapped dust density reaches about $30\%$ of the gas density \citep[see also][]{Fu2014,Raettig2015}.
Here, we find that if we initiate the simulation above that value, the vortex does not even develop.
Even if we decrease the ring width considerably to $\sigma\!=\!H$, the instability remains suppressed by the dust.

However, an alternative is that the gas is secondary but dominated by species other than CO.  
In secondary gas scenarios the instantaneous CO mass is set by the volatile mix of the parent bodies and how fast CO photodissociates into C and O.  
Unshielded CO around debris discs typically survives only $\sim\!100$–$200$\,yr \citep[often quoted as $\approx120$\,yr for the interstellar radiation field; e.g.][]{Matra2015, Marino2016, Kral2017}, so most of the total gas mass may reside in atomic carbon and oxygen unless CO is shielded in which case CO tends to dominate the gas mass in secondary models \citep{Kral2019, Marino2020, Marino2022}. This is because CO shielding by carbon can dramatically prolong the CO lifetime. Once CO becomes shielded, its photodissociation rate can drop by orders of magnitude.
In such shielded discs, the carbon abundance is regulated: higher carbon abundance increases shielding and thus carbon and oxygen production from CO is diminished. Whether this applies to \HD{} remains open.

For \HD{}, the currently reported C mass is
$M_{\rm C\,I}\!\approx\!4\!\times\!10^{-3}\,M_\oplus$, corresponding to a column density of $N\!\sim\!2\!\times\!10^{17}\,{\rm cm}^{-2}$\citep[][]{Cataldi2023}.
In other words, given the observational estimates at the time of writing, the carbon mass in \HD{} is significantly lower than the CO mass and close to the value where shielding starts to become efficient \citep[$\gtrsim\!10^{-17}\,{\rm cm}^{-2}$,][]{Kral2019, Marino2020, Marino2022}.
Still, we carried out two additional simulations with total gas masses of $2.5\,M_\oplus$ and $5\,M_\oplus$, adopting dust-to-gas mass ratios of $f_{\rm dtg}\!=\!0.084$ and $f_{\rm dtg}\!=\!0.042$, respectively.
These runs are meant to represent cases in which CO is abundant but does not dominate the overall gas mass budget.

\begin{figure*}
    \sidecaption
    \includegraphics[width=12cm]{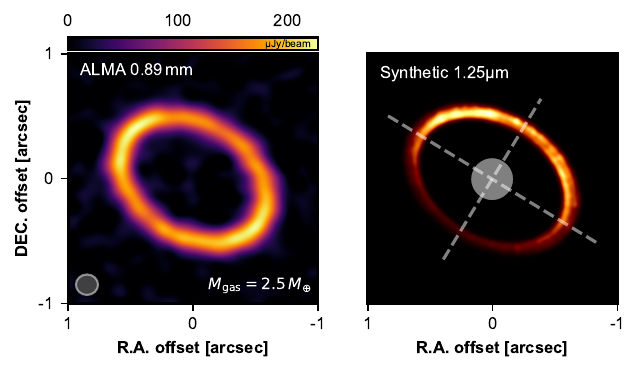}
    \caption{
    Synthetic images for the $M_{\rm gas}\!=\!2.5\,M_\oplus$ model:
    ALMA band\,7 continuum (left) and $J$-band polarised intensity (right).
    Dashed lines in the right panel mark the projected semi-major and semi-minor axes, according to the employed PA of $58.7^\circ$ \citep{overview_arks}.
    }
    \label{fig:lowgasmass}
\end{figure*}

In this simulation the instability does develop in the gas, but the largest dust grains ($a\!\gtrsim\! 3\,\mathrm{mm}$) have ${\rm St}\!\gg\! 1$ and are therefore only weakly affected by gas drag.
Owing to the combination of low gas surface density and the resulting high ${\rm St}$, no clear azimuthal asymmetry appears in the mm-sized grains that dominate the band\,7 continuum.
By contrast, small grains of $\sim 10\,\mu\mathrm{m}$ decouple from the gas and do develop an azimuthal
asymmetry.
Consequently, this model fails to reproduce the observed combination in \HD{} of an asymmetric ALMA continuum arc
and a symmetric SPHERE scattered-light image (Fig.~\ref{fig:lowgasmass}).

\begin{figure*}
    \sidecaption
    \includegraphics[width=12cm]{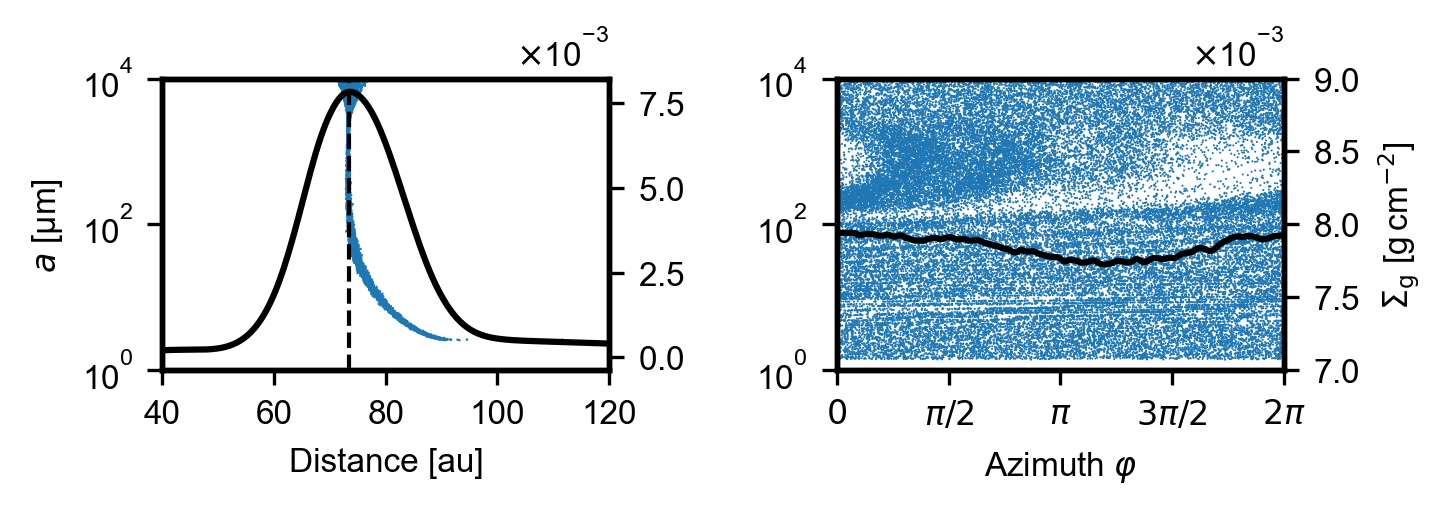}
    \caption{Particle distributions for the $M_{\rm gas}\!=\!5\,M_\oplus$ model after 150 orbits at $r_0$.
    Left: radial profiles (cf. Fig.~\ref{fig:rad_profile});
    Right: azimuthal profiles (cf. Fig.~\ref{fig:azi_profile}).
    In both panels the black curve shows the gas surface density (right-hand axis).}
    \label{fig:lowmass_profiles}
\end{figure*}

The $5\,M_\oplus$ case behaves differently. After the gas asymmetry weakens, only a specific grain-size window, $100\,\mu\mathrm{m}\!\lesssim a \lesssim\!3\,\mathrm{mm}$, retains a strong azimuthal contrast (Fig.~\ref{fig:lowmass_profiles}).
In the radial dimension, $a\!\lesssim\!50\,\mu\mathrm{m}$ grains are gradually pushed outward by the combined action of radiation pressure and gas drag; particles with $50\,\mu\mathrm{m}\!\lesssim a\!\lesssim\!2\,\mathrm{mm}$ are efficiently confined at the pressure maximum; and $a\!\gtrsim\!3\,\mathrm{mm}$ grains, with ${\rm St}\!\gg\!1$, remain largely on their
initial orbits.
Thus, the azimuthal asymmetry is carried primarily by those grains that are radially trapped (and hence concentrated) but not so large as to be entirely decoupled.

\begin{figure*}
    \centering
    \includegraphics[width=0.80\textwidth]{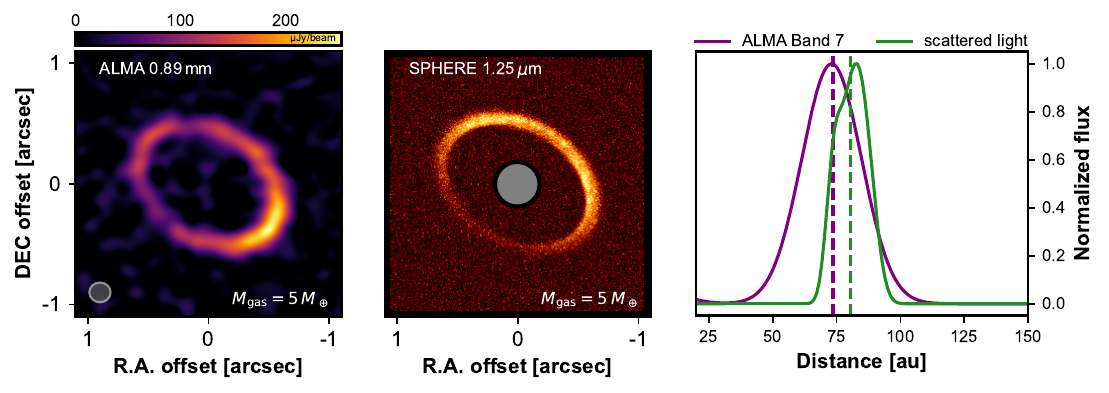}
    \caption{
    Synthetic observables for the $M_{\rm gas}\!=\!5\,M_\oplus$ model. Left: ALMA band\,7 continuum; Centre: SPHERE $J$-band polarised intensity; Right: deprojected radial profiles of the two rings.
    }
    \label{fig:triple_5Mearth}
\end{figure*}

Fig.~\ref{fig:triple_5Mearth} presents the corresponding synthetic observations: ALMA band\,7 (left panel), SPHERE $J$-band polarised intensity (central panel), and deprojected radial profiles of both (right panel).
The model qualitatively reproduces the main features seen in \HD{}: an asymmetric mm continuum ring coupled with a more uniform scattered-light ring, with a fidelity comparable to the fiducial ($M_{\rm gas}\!=\!50\,M_\oplus$) model. The radial offset between the rings in the synthetic ALMA and SPHERE images is 10.9\%, in good agreement with the observations, though slightly larger.

\section{Discussion}\label{sec:discussion}

\subsection{Hybrid rather than debris disc}\label{subsec:hybrid}

Protoplanetary discs have been detected around stars as old as 23\,Myrs \citep[V4046\,Sgr,][]{Mamajek2014}. 
Considering \HD{}'s estimated age of 16\,Myrs, evidence for a primordial origin of the detected gas would question its classification as a traditional debris disc.
Instead, it might represent a rather old, dispersing protoplanetary disc in a transitional phase.
Unfortunately, the term “transition disc” has historically been connoted with discs that host large inner cavities that were classified through dust continuum deficits at near- or mid-infrared wavelengths (e.g. \citealt{Strom1989,Espaillat2014}).
However, it has since been found that most of these so-called transition discs are actually steady, persistent environments, contradicting the notion of an imminent transition.
To differentiate from the classical connotation of transition discs, we will here use the term “hybrid disc”, following \citet{Kospal2013}, for a system that mainly retains its  gas from the protoplanetary disc while its solids are at debris-like levels (although not necessarily collisionally generated as in standard debris disc definitions).

\citet{Moor2017} surveyed several young A-type stars and detected CO in 11 out of 16 young ($\leq \!42\,$Myr) classified debris discs, suggesting that they might indeed be such hybrid discs, a sub-population of late-stage discs around intermediate-mass stars that still harbor significant gas.
As argued in Sec.~\ref{subsec:lowlimit}, the observations are incompatible with the gas drag scenario for total gas masses of $M_{\rm gas}\!\lesssim\! 2.5\,M_\oplus$.
Consequently, a confirmation of the gas drag shaping the continuum emission in \HD{} would strongly point to the hybrid disc scenario.
In this case, \HD{} might be undergoing final dispersal through mechanisms such as photoevaporation \citep[e.g.][]{Nakatani2021,Nakatani2023}.

Under photoevaporation, high-energy stellar photons heat the gas to sufficiently high temperatures to drive mass loss from the disc (e.g., \citealt{Owen2012,Ercolano2018}).
As gas is removed from the inside out, the outer ring may become progressively “steepened” or radially compressed, while radiation pressure and stellar winds further erode any remaining material.
Thus, the presence of ${\sim} 0.15\,M_{\oplus}$ of CO in \HD{} could reflect late stage gas dispersal rather than a purely secondary gas origin (i.e., collisionally generated, see also \citealt{Kral2017} for secondary gas origins).
If so, \HD{} may well illustrate how a disc in the final stages of protoplanetary evolution can masquerade as a debris disc while retaining enough gas to affect dust dynamics and reveal ongoing gas dispersal. The co-location of dust and gas in this case could be simply explained via dust-gas interactions that drive the large dust grains towards the pressure maximum.

In this picture, gas retention beyond $\!\sim\!10\,\mathrm{Myr}$ implies that photoevaporation and disc winds operate inefficiently, as expected when the stellar high-energy drivers are intrinsically weak. This naturally explains the prevalence of gas detections around A-type stars \citep[][]{Moor2017}, implying low X-ray and extreme-UV output \citep[e.g.][]{Nakatani2023}. Moreover, if very small grains and Polycyclic Aromatic Hydrocarbons are depleted, the efficiency of far-UV (FUV) photoelectric heating drops sharply, further suppressing FUV-driven photoevaporation \citep[e.g.][]{Nakatani2021,Coleman2022,Nakatani2023,OOyama2025}. Thus, the classification of \HD{} as an A-type star and the proposed lack of grains with $a\!\lesssim\!1\mu{\rm m}$ \citep[][]{Perrot2023} support the notion of slow protoplanetary disc dispersal and thereby the primordial gas scenario.

\subsection{Observational tests for trapping by gas drag}
\label{subsec:observe_predictions}

\subsubsection{$50\,M_\oplus$ versus $5\,M_\oplus$}
The three key observables in \HD{} are (i) a confined arc in the ALMA band~7 continuum, (ii) a $J$-band morphology dominated by the scattering phase function, and (iii) a radial offset between the sub-mm and NIR rings.
They are reproduced by both our fiducial model with $M_{\rm gas}\!=\!50\,M_\oplus$ and the lower-mass model with $M_{\rm gas}\!=\!5\,M_\oplus$.
The hydrodynamical simulations, however, reveal a crucial difference: the size-dependence of the azimuthal structure near the arc.
In the low-mass case, the largest grains remain largely unaffected by gas drag and show little or no azimuthal trapping.
Because different ALMA bands probe different characteristic grain sizes, multi-band continuum imaging offers a way to break this degeneracy in $M_{\rm gas}$.

To test this, we perform additional continuum radiative transfer at 0.45, 1.25, and 2.2\,mm, representative of ALMA bands~9, 6, and 4, respectively.
From each image, we deproject the ring and extract an azimuthal profile by integrating the flux over a narrow radial window ($\pm 2$\,au) centred on the radial peak.
We fit the deprojected profile by a Gaussian function atop a constant background (representing the azimuthally symmetric ring component), and normalise the profile by the background.

\begin{figure*}
    \sidecaption
    \includegraphics[width=12cm]{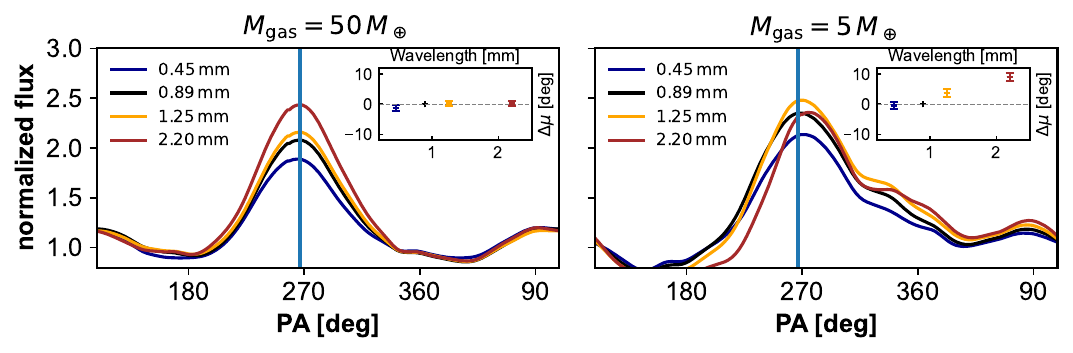}
    \caption{Normalised azimuthal brightness profiles of the deprojected ring at four wavelengths, 0.45, 0.89, 1.25, and 2.20\,mm (ALMA bands 9, 7, 6, and 4), as a function of position angle (PA). PA increases east of north, in the direction of the system’s orbital motion. Left: model with \(M_{\rm gas}\!=\!50\,M_\oplus\). Right: model with \(M_{\rm gas}\!=\!5\,M_\oplus\). Profiles are normalised to the azimuthally smooth ring component (unity corresponds to the smooth-ring level). Insets show the offset of the peak PA at each wavelength relative to the band\,7 peak measured in \citet{hd121617_arks}. In the high-mass model, the peak PA is nearly wavelength-independent and the contrast increases with wavelength; in the low-mass model, the contrast drops at 2.2\,mm and the peak shifts systematically toward larger PA with increasing wavelength.}
    \label{fig:multiband_az_profile}
\end{figure*}
Figure~\ref{fig:multiband_az_profile} compares the azimuthal profiles for the two gas masses.
We focus on two diagnostics: (i) the contrast of the intensity maximum relative to the smooth (axisymmetric) component, and (ii) the azimuth of the intensity maximum, obtained by fitting a Gaussian to each profile. 

For $M_{\rm gas}\!=\!50\,M_\oplus$ (left panel), the peak-to-background contrast at 0.89\,mm is $2.08\pm0.02$, the PA of the peak location is $28.0^\circ \!\pm\! 0.2^\circ$ and the azimuthal FWHM of the arc is $67^\circ \pm 1^\circ$.
In the multi-band analysis we see that the contrast increases systematically with wavelength from band~9 to band~4, while the azimuth of the maximum remains constant within the margin of error.
In the $M_{\rm gas}\!=\!5\,M_\oplus$ model (right panel), the peak-to-background contrast at 0.89\,mm is $2.35\pm0.03$, the PA of the peak location is $27.9^\circ \!\pm\! 0.2^\circ$ and the azimuthal FWHM of the arc is $87^\circ \! \pm \! 2^\circ$.
In the multi-wavelength analysis, both diagnostics behave differently: the contrast rises from band~9 to band~6 but then declines towards band~4, and the peak shifts in azimuth by $3.7^\circ\!\pm\!1.3^\circ$ (band~6) and $9.1^\circ\!\pm\!1.4^\circ$ (band~4) relative to the observed band~7 maximum \citep[][]{hd121617_arks}. 

The declining contrast between bands~6 and 4 in the low-mass case arises because the largest grains, which contribute more to the longer-wavelength emission, are azimuthally more symmetric when the gas mass is lower.
As the observing wavelength increases, the measurement becomes more sensitive to this nearly axisymmetric population, diluting the arc contrast.
The systematic azimuthal shift is likewise a consequence of size-dependent dust–gas coupling: grains with longer stopping times respond differently to gas asymmetries and tend to lead smaller grains in azimuth \citep{Baruteau2019}. 

In summary, a monotonic increase of arc contrast with wavelength and a stationary arc azimuth favour the higher gas mass, whereas a non-monotonic contrast (peaking near band~6) together with a wavelength-dependent azimuthal shift point to the lower-mass scenario.

\subsubsection{Planet versus gas drag}
While our hydrodynamical simulations demonstrate that the azimuthally asymmetric mm ring observed in \HD{} can be reproduced through gas–dust interactions, this does not provide a conclusive explanation for such a phenomenon.

An alternative explanation is that the disc asymmetry arises from interactions with an unseen planet (Pearce et al. in prep.).
In this scenario, a planet located interior to the debris disc is (artificially) forced to undergo outward migration, similar to Neptune's planetesimal-driven migration in the Solar System \citep{Malhotra1993, Hahn1999}.
As the planet migrates, its mean-motion resonances sweep through the disc, capturing debris bodies in resonances and creating asymmetric clumps in their distribution.
The structure, location, and number of these clumps depend on the specific resonances in which the debris becomes trapped, which is a probabilistic event depending on the planetary migration rate \citep{Wyatt2003}.
This simple migration model has previously been proposed to explain a clump of planetesimals around $\beta$\,Pic \citep{Matra2019}, an asymmetric disc around q$^1$\,Eri \citep{Lovell2021}, and tentative clumps in the debris disc around $\epsilon$ Eri \citep{Booth2023}.
In this purely planetary explanation, it remains to be seen how the measured content of CO gas would impact the dust distribution.
Additionally, collective gravity of debris particles may play an important dynamical role by hindering the planet's migration  \citep{Hahn1999}, detuning the resonances \citep{Murray2022}, and/or suppressing the debris eccentricities \citep{Sefilian2024}.
Consequently, further tests are needed to distinguish between these two possibilities.

One test to differentiate between the planetary and gas drag scenario involves observations in the time domain.
As shown by \citet{Wyatt2006}, planetesimals in an outward-migrating planetary system become predominantly trapped in 2:1 or 3:2 mean-motion resonances.
This implies a planetary orbit at $r_{2:1} \!=\! 57\,\mathrm{au}$ or $r_{3:2} \!=\! 47\,\mathrm{au}$.
If the arc is sculpted by mean-motion resonances, it should orbit with the planet’s own period, rather than the local Keplerian period.
At the ring's radius, the Keplerian orbital period is approximately 474\,years, so if the dust arc is induced by gas drag it will complete one full rotation on that timescale.
However, if the ring was in a 2:1 resonance with an interior planet, it would orbit in 237\,years; in a 3:2 resonance, 316\,years.
Future measurements of the arc's azimuthal shift can be compared against these predictions to distinguish between the two scenarios.
If we require a minimum shift of half the beam size \citep[$\sim\! 7\,\mathrm{au}$,][]{overview_arks} to confidently register an azimuthal displacement, then for the 2:1, 3:2, and gas-dominated scenarios the necessary intervals are $\Delta t_{2:1} \!=\! 4.6\,\mathrm{yr}$, $\Delta t_{3:2} \!=\! 6.1\,\mathrm{yr}$, and $\Delta t_{\mathrm{gas}} \!=\! 9.2\,\mathrm{yr}$, respectively. 
Given that the initial dataset was obtained between October 2022 and May 2023, a second observational epoch in or after 2033 would detect a positional shift of the arc even under local Keplerian motion.
An observation in 2030 could already confirm or rule out the scenario of resonant trapping.

\subsection{Caveats and prospects}
This study demonstrates a new avenue for simulating debris disc dynamics by adapting numerical techniques originally developed for protoplanetary discs.
However, several important considerations remain unaddressed and invite further investigation.
An obvious question is about the origin of the radially confined ring, which we imposed as an arbitrary initial condition, assuming that the ring is an inherited structure from the protoplanetary disc phase.
Indeed, \citet{Orcajo2025} found that at the latest stage of a surveyed protoplanetary disc sample, the dust typically converges to a narrow isolated ring.
One plausible explanation for the inner cavity and the concentration of material at its edge is the presence of a sufficiently massive planet.
In fact, in protoplanetary discs, planet–disc interactions have previously been shown to induce vortices at the edge of the gap, similar to what has been observed for \HD{} \citep[e.g.][]{devalBorro2007}. For HD~121617, only planets more massive than $7\,M_{\rm Jup}$ are ruled-out at the inner edge of the disc \citep{scat_arks}.
Even in the absence of planets, similar structures can be induced by steep transitions of the viscosity profile, as expected between active and passive turbulent zones \citep{Flock2015}.
Future work should therefore focus on examining debris disc hydrodynamics under the influence of planetary perturbations, aiming to capture both the formation and evolution of these rings in a more comprehensive manner.

Another important flaw is the omission of dust growth, fragmentation, and production by planetesmals (if present) in the numerical evolution of the ring.
In debris discs, dust is continuously generated by collisions between larger bodies and may be destroyed in collisions between smaller particles.
This ongoing process may introduce a constant redistribution of particle sizes and reduce the size-dependence of the overall density distribution.
The collisional timescale can be shorter than the timescale on which gas drag imposes a spatial size seggregation in the disc, depending on the gas density and other parameters. This is discussed in the context of another gas-bearing debris disc, HD\,131835, in a companion ARKS study \citep{hd131835_arks}.
In a study dedicated primarily to size-dependent radial trapping in debris discs, Olofsson et al. (in prep.) carry out a parameter sweep and find that the SPHERE–ALMA offset is governed primarily by the competition between gas mass and optical depth (a proxy for the collisional timescale). They find that there is a “sweet spot” in dust-to-gas ratio where the offset is maximised. They find that at high gas mass and low optical depth, a secondary mm ring can appear and mimic a gap carved by a planet.

However, the high collision rates typically assumed in debris discs rely on large relative velocities driven by orbital eccentricities and inclinations.
In the presence of substantial gas, these orbital elements are rapidly damped unless there is a continuous mechanism re-exciting them, thereby lowering the collisional velocities compared to a gas-free scenario.
Hence, if the gas is primordial, its damping of relative velocities would suppress destructive collisions and increase the survival time of solids. This raises the possibility that the dust population is at least partly primordial rather than maintained solely by planetesimal collisions.
For a detailed consideration of dust trapped inside a gas ring and the relevance of collisions in the context of a debris disc, we further refer to the discussion Section~4.1 of \citet{Pearce2020}.

Gas drag might also play an important role in other systems that are classified as debris discs.
A prominent example is the exoKuiper belt around HD\,181327, where CO is co-located with a band\,6 continuum ring \citep{Marino2016}.
In this system, the observational evidence reveals notable differences: the millimetre continuum appears comparatively smooth, whereas optical scattered-light observations with the \emph{Hubble Space Telescope} show an azimuthal asymmetry \citep{Stark2014,Fox2025}.
This trend is qualitatively consistent with a reduced gas content compared to \HD{} (as illustrated, for example, in Fig.~\ref{fig:lowgasmass}) and is in agreement with the low CO mass inferred for HD\,181327 ($M_{\rm CO} \!\sim\! 2\times 10^{-6}\,M_\oplus$; \citealt{Marino2016}).
A definitive test of this interpretation would require dedicated, system-specific simulations, which are beyond the scope of this work.

\section{Conclusions}
\label{sec:conclusions}

We used hydrodynamical simulations with self-consistent dust-gas coupling to test whether gas drag can explain the submillimetre continuum arc observed in the debris disc of \HD{}. Our main findings can be summarised as follows:

\begin{enumerate}
    \item {Gas drag trapping can reproduce the ALMA arc:}
    A marginally stable gas ring naturally produces azimuthal dust concentrations when the aerodynamic coupling is sufficiently high. In particular, models with total gas masses of $M_{\rm gas}\!=\!50\,M_\oplus$ and $5\,M_\oplus$ both reproduce the observed ALMA band\,7 continuum arc.

    \item {Size-dependent dynamics and radiation pressure explain NIR/mm differences:}
    Stellar radiation pressure from the A-type host causes small grains to orbit sub-Keplerian and drift outward, while millimetre-sized particles concentrate at the pressure maximum. As a result, $\mu$m grains traced in scattered light are displaced outward relative to the mm continuum ring, whereas mm grains form the azimuthal arc. The observed offset between the near-infrared and sub-mm rings thus arises naturally from the different responses of grain sizes to radiation pressure and gas drag.

 \item {Constraints on the gas mass:}
    The outward shift of the scattered-light ring requires that $\mu$m grains are not perfectly entrained, implying a limit on the coupling and therefore on the total gas content. From our models we infer an upper limit of $\sim\!250\,M_\oplus$ for the total gas mass; conversely, the presence of a pronounced sub-mm arc and the absence of scattered-light asymmetries with respect to the semi-minor axis require at least a few Earth masses of gas, with a conservative lower bound of $\sim\!2.5\,M_\oplus$. Together, these limits suggest that \HD{} may contain more gas than its estimated CO content and that predicted from secondary models, favouring a hybrid-disc (primordial gas) scenario.

 \item {Observational tests:}
    The simulations predict a wavelength-dependent azimuthal contrast because only a restricted range of grain sizes is efficiently azimuthally confined. In the $5\,M_\oplus$ model, the largest grains remain closer to an azimuthally uniform distribution than in the $50\,M_\oplus$ model, reducing the contrast at longer wavelengths. Moreover, the azimuthal overdensity should orbit at the local Keplerian rate. Multi-wavelength ALMA imaging that probes different grain sizes and a second band\,7 epoch to track the pattern’s orbital motion would provide decisive tests of gas drag trapping versus alternatives such as resonant confinement by an outward-migrating interior planet.

\end{enumerate}

\noindent We note that the quoted gas mass bounds are model-dependent. They are expected to vary with assumptions about the turbulent viscosity, gas temperature and scale height, ring width, and the dust size distribution and opacities that set the relative contributions of trapped versus background grains.

\medskip
\noindent More broadly, our results show that hydrodynamical modelling of continuum substructures can place meaningful constraints on the hidden gas and its origin in debris discs. For \HD{}, a hybrid-disc interpretation can resolve apparent spatial differences between the emission profiles of CO gas and dust structures, yet it implies a more complex evolutionary pathway bridging the protoplanetary and debris disc phases. Future work that couples dust production and destruction with planet-disc interactions will be crucial to establish whether \HD{} and analogous systems trace the final stages of protoplanetary disc dispersal or a predominantly secondary gas origin.

\section*{Data availability}
The ARKS data used in this paper and others can be found in the \href{https://dataverse.harvard.edu/dataverse/arkslp}{ARKS dataverse}.

\begin{acknowledgements}
PW and SP acknowledge support from FONDECYT grants 3220399 and 1231663 and ANID -- Millennium Science Initiative Program -- Center Code NCN2024\_001. SM acknowledges funding by the Royal Society through a Royal Society University Research Fellowship (URF-R1-221669) and the European Union through the FEED ERC project (grant number 101162711). MRJ acknowledges support from the European Union's Horizon Europe Programme under the Marie Sklodowska-Curie grant agreement no. 101064124 and funding provided by the Institute of Physics Belgrade, through the grant by the Ministry of Science, Technological Development, and Innovations of the Republic of Serbia. TDP is supported by a UKRI Stephen Hawking Fellowship and a Warwick Prize Fellowship, the latter made possible by a generous philanthropic donation. A.A.S. is supported by the Heising-Simons Foundation through a 51 Pegasi b Fellowship. JBL acknowledges the Smithsonian Institute for funding via a Submillimeter Array (SMA) Fellowship, and the North American ALMA Science Center (NAASC) for funding via an ALMA Ambassadorship. CdB acknowledges support from the Spanish Ministerio de Ciencia, Innovaci\'on y Universidades (MICIU) and the European Regional Development Fund (ERDF) under reference PID2023-153342NB-I00/10.13039/501100011033, from the Beatriz Galindo Senior Fellowship BG22/00166 funded by the MICIU, and the support from the Universidad de La Laguna (ULL) and the Consejer\'ia de Econom\'ia, Conocimiento y Empleo of the Gobierno de Canarias. AMH acknowledges support from the National Science Foundation under Grant No. AST-2307920. SMM acknowledges funding by the European Union through the E-BEANS ERC project (grant number 100117693), and by the Irish research Council (IRC) under grant number IRCLA- 2022-3788. Views and opinions expressed are however those of the author(s) only and do not necessarily reflect those of the European Union or the European Research Council Executive Agency. Neither the European Union nor the granting authority can be held responsible for them. AB acknowledges research support by the Irish Research Council under grant GOIPG/2022/1895. LM acknowledges funding by the European Union through the E-BEANS ERC project (grant number 100117693), and by the Irish research Council (IRC) under grant number IRCLA- 2022-3788. Views and opinions expressed are however those of the author(s) only and do not necessarily reflect those of the European Union or the European Research Council Executive Agency. Neither the European Union nor the granting authority can be held responsible for them. JM acknowledges funding from the Agence Nationale de la Recherche through the DDISK project (grant No. ANR-21-CE31-0015) and from the PNP (French National Planetology Program) through the EPOPEE project. Support for BZ was provided by The Brinson Foundation. EC acknowledges support from NASA STScI grant HST-AR-16608.001-A and the Simons Foundation. MB acknowledges funding from the Agence Nationale de la Recherche through the DDISK project (grant No. ANR-21-CE31-0015). This work was also supported by the NKFIH NKKP grant ADVANCED 149943 and the NKFIH excellence grant TKP2021-NKTA-64. Project no.149943 has been implemented with the support provided by the Ministry of Culture and Innovation of Hungary from the National Research, Development and Innovation Fund, financed under the NKKP ADVANCED funding scheme. This research was partially supported by the supercomputing infrastructure of the NLHPC (CCSS210001). This work made use of the Puelche cluster hosted at CIRAS/USACH.
This paper makes use of the following ALMA data: ADS/JAO.ALMA\# 2022.1.00338.L, 2012.1.00142.S, 2012.1.00198.S, 2015.1.01260.S, 2016.1.00104.S, 2016.1.00195.S, 2016.1.00907.S, 2017.1.00167.S, 2017.1.00825.S, 2018.1.01222.S and 2019.1.00189.S. ALMA is a partnership of ESO (representing its member states), NSF (USA) and NINS (Japan), together with NRC (Canada), MOST and ASIAA (Taiwan), and KASI (Republic of Korea), in cooperation with the Republic of Chile. The Joint ALMA Observatory is operated by ESO, AUI/NRAO and NAOJ. The National Radio Astronomy Observatory is a facility of the National Science Foundation operated under cooperative agreement by Associated Universities, Inc. The project leading to this publication has received support from ORP, that is funded by the European Union’s Horizon 2020 research and innovation programme under grant agreement No 101004719 [ORP]. We are grateful for the help of the UK node of the European ARC in answering our questions and producing calibrated measurement sets. This research used the Canadian Advanced Network For Astronomy Research (CANFAR) operated in partnership by the Canadian Astronomy Data Centre and The Digital Research Alliance of Canada with support from the National Research Council of Canada the Canadian Space Agency, CANARIE and the Canadian Foundation for Innovation.
\end{acknowledgements}

\bibliography{references}{}
\bibliographystyle{aa}

\begin{appendix}
\section{Validation of momentum feedback from a size-distribution}\label{appendix:feedback}

Accurate modelling of dust-to-gas momentum feedback is crucial when the dust-to-gas mass ratio is large. In such cases, the simulation outcome strongly depends on a sufficiently precise implementation of the drag forces exerted by the dust on the gas. Since this aspect has not yet been tested in Dusty FARGO-ADSG, we performed a simple drift test on a one-dimensional grid.

We initialised the gas surface density with a radial power-law:
\begin{equation}
    \Sigma_{\mathrm{g}}(r) = \Sigma_{\mathrm{g},0}\,\left(\frac{r}{r_0}\right)^{-0.5}.
\end{equation}
We also set the dust density, sampled by the particles, to follow the same power-law profile, ensuring an initial constant dust-to-gas ratio across all radii.
We considered an inviscid scenario and discard stochastic kicks between particles.
No exact analytical solution exists for dust and gas velocities under a continuous size distribution, only for multiple discrete species \citep{Benitez2019}. We therefore compared our numerical results to the analytical expressions for radial and azimuthal velocities derived for discrete sizes \citep[][their Eqs.\,70--73]{Benitez2019}.
Instead of randomly sampling sizes from a continuous distribution, we split the dust into a small number of discrete bins.
In all the tests, we used the cloud-in-cell interpolation scheme.
For improved display of the results, we binned every ten simulated particles to a representative data point.

\subsection{Single-sized grains}

\begin{figure}
    \centering
    \includegraphics[width=\columnwidth]{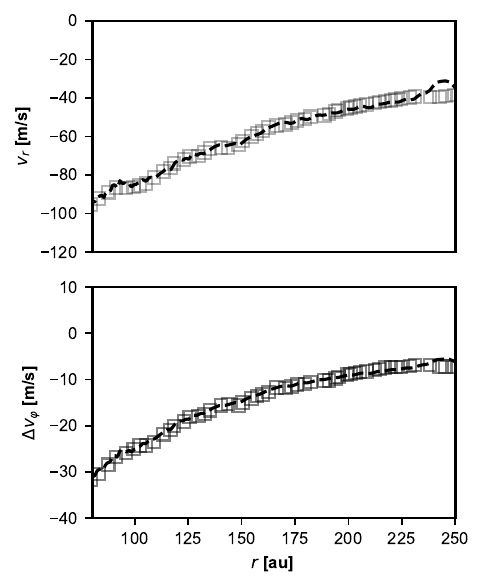}
    \caption{Validation of the dust feedback implementation for single-sized grains ($a=0.1\,$mm, $\Sigma_{\rm g,0}=0.15\,{\rm g}\,{\rm cm}^{-2}$, $f_{\rm dtg}=0.1$).
    \emph{Top:} Radial velocity; \emph{Bottom:} Difference between azimuthal velocity and local Keplerian velocity. The dashed lines are the analytical solutions \citep[][Eqs.\,72--73]{Benitez2019}, and the squares represent the dust particle velocities. For clarity, we binned the velocity of ten neighbouring particles to one representative data point.}
    \label{fig:test_1size}
\end{figure}

In an initial test, we set $\Sigma_{\mathrm{g},0} \!=\! 0.15\,\mathrm{g\,cm}^{-2}$ and distributed $10^3$ particles of the same size, $a\!=\!0.1\,\mathrm{mm}$, over 512 cells.
This configuration corresponded to a dust-to-gas ratio of $0.1$ and the particles Stokes number corresponded to ${\rm St}\!\sim\!0.1-1$, depending on the location in the disc.
Both dust and gas started with zero radial velocity.
The gas azimuthal velocity was set to its pressure-supported sub-Keplerian profile (Eq.~\ref{equ:vphi}), the dust particles' azimuthal velocity was set to Keplerian.
We find that the particles’ velocities rapidly converge to the analytical predictions.
Figure~\ref{fig:test_1size} shows the radial velocities and azimuthal deviation from Keplerian velocities after five orbits at $r_0=75\,$au.

\subsection{Multi-sized grains}

Next, we considered two discrete dust sizes in the same setup, $a\!=\!0.1\,\mathrm{mm}$ and $a\!=\!1\,\mathrm{mm}$, using $10^4$ total particles. We examined two scenarios, first with $\Sigma_{\mathrm{g},0} \!=\! 0.15\,\mathrm{g\,cm}^{-2}$ and $f_{\rm dtg}\!=\!0.1$, then $\Sigma_{\mathrm{g},0} \!=\! 0.015\,\mathrm{g\,cm}^{-2}$ and $f_{\rm dtg}\!=\!1$.

\begin{figure*}
    \centering
    \includegraphics[width=\textwidth]{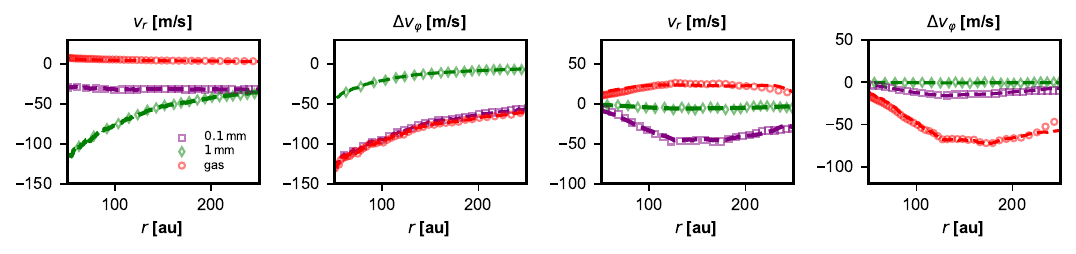}
    \caption{Drift test with two discrete dust sizes. \emph{Left panels:} Radial velocity and azimuthal deviation for $f_{\rm dtg}\!=\!0.1$. \emph{Right panels:} Same, but with the gas density lowered by a factor of 10, resulting in $f_{\rm dtg}\!=\!1$. The dashed lines show the analytical solutions.}
    \label{fig:fb_2sizes}
\end{figure*}

\begin{figure}
    \centering
    \includegraphics[width=\columnwidth]{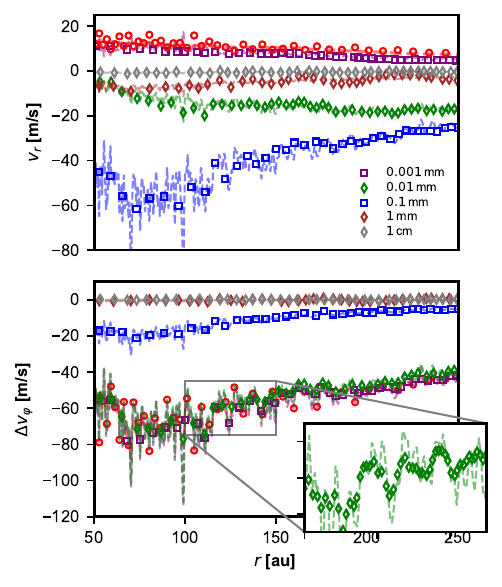}
    \caption{Same as Fig.~\ref{fig:fb_2sizes}, but with 5 discrete sizes. The total dust-to-gas ratio was unity, and each dust-size bin carried an equal fraction of the total dust mass.}
    \label{fig:fb_5sizes}
\end{figure}

Figure~\ref{fig:fb_2sizes} displays the resulting radial velocities and azimuthal deviations from Keplerian rotation for both dust species and the gas, in both the $f_{\rm dtg}\!=\!0.1$ and $f_{\rm dtg}\!=\!1$ cases. We further tested five discrete sizes, each contributing the same dust mass, to achieve an overall dust-to-gas ratio of unity (see Figure~\ref{fig:fb_5sizes}).
In all those tests, the analytical velocity profiles are well matched by the numerical solution.

\subsection{Continuous size distribution}

In the simulations presented in this study, we drew particle sizes from a continuous dust size distribution.
For a given dust size distribution, dust-to-gas ratio, and gas density profile, we can derive semi-analytic solutions for the corresponding gas radial and azimuthal velocities.
Specifically, we discretised the continuous distribution into a large number of bins ($N_{\rm dust} \!=\! 1000$) and numerically solve Eqs.\,70--71 of \citet{Benitez2019}.
We then compare these solutions with the hydrodynamical results from {Dusty FARGO-ADSG} at ten orbits.
Because each dust size exhibits a different drift behaviour, the local dust-to-gas ratio and size distribution begin to deviate from their initial profiles after only a few orbits, so limiting the evolution to ten orbits ensured closer adherence to the initial conditions, while giving enough time for the numerical solution (markers) to approach the semi-analytical solution (lines).

\begin{figure}
    \centering
    \includegraphics[width=\columnwidth]{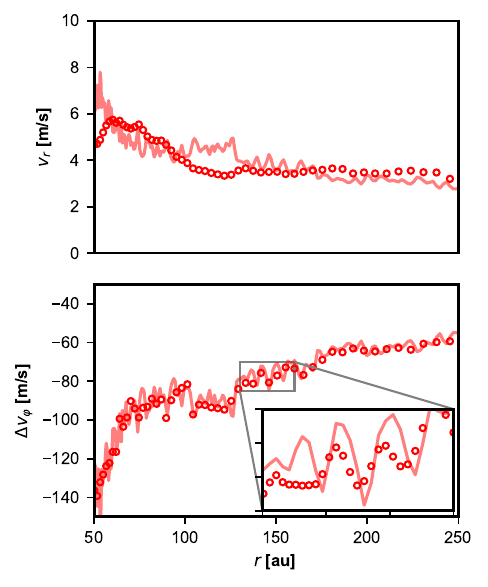}
    \caption{Gas velocities for a continuous dust size distribution with a power-law index of $-3.5$, ranging from $0.1\,\mu\mathrm{m}$ to $1\,\mathrm{cm}$ at a dust-to-gas ratio of $f_{\rm dtg} \!=\! 0.1$. The solid lines show the semi-analytic solutions from Eqs.\,70 (top panel) and 71 (bottom panel) in \citet{Benitez2019}, based on the gas density profile after ten orbits. Circles indicate the corresponding gas velocities computed by {Dusty FARGO-ADSG} at the same output time.}
    \label{fig:test_continuous}
\end{figure}

Figure~\ref{fig:test_continuous} illustrates this comparison for $f_{\rm dtg} \!=\! 0.2$. Overall, there is good agreement between the two approaches. Any differences are not necessarily due to numerical inaccuracies, but may instead arise from gradual departures in the actual dust-to-gas ratio and size distribution from their assumed initial state in the semi-analytic model.

\section{Evolution of the asymmetry in the gas ring}\label{appendix:timeseries}
The initial condition of our setup is unstable to the Rossby-Wave Instability.
This instability generates the asymmetric gas structure that is then imparted onto the solids via friction.
However, the instability leads to a radial spreading of the dust ring and, without external forcing, the asymmetry dissipates with time.
\begin{figure*}
    \centering
    \includegraphics[width=\textwidth]{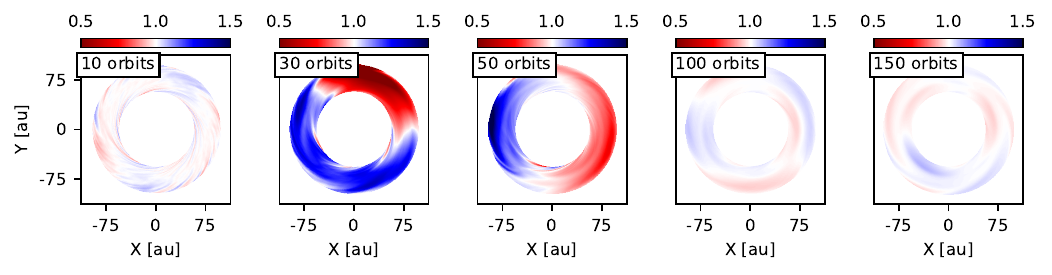}
    \caption{Evolution of the asymmetry generated by the unstable gas ring. The colour shows the gas surface density divided by the azimuthally averaged gas surface density. The displayed outputs correspond to $4.7\,{\rm kyr}$, $14.1\,{\rm kyr}$, $23.6\,{\rm kyr}$, $47\,{\rm kyr}$ and $71\,{\rm kyr}$, respectively.}
    \label{fig:timeseries}
\end{figure*}

Fig.~\ref{fig:timeseries} presents the evolution of the azimuthal asymmetry.
For each snapshot we divided the gas surface density by its azimuthally averaged profile.
In the figure, blue marks regions denser than the local mean, and red marks depleted sectors of the ring.
We display this deviation only for regions where the local mean is larger than $0.01\,{\rm g}\,{\rm cm}^{-2}$ (corresponding to $\sim\!7\%$ of $\Sigma_0$) to not be confused by contributions from low-density regions.
The instability generates its strongest contrast after about 25 orbits.
By the time the fiducial model is analysed, after 150 orbits, the asymmetry in the gas has already weakened significantly.

\FloatBarrier
\twocolumn

\FloatBarrier
\clearpage

\end{appendix}
\end{document}